\documentclass[manuscript]{aastex}
\usepackage{lscape}
\shorttitle{The AKARI Catalog}
\shortauthors{Kato et al.}

\begin{document}%
%\Received{}
%\Accepted{}
%\Published{}
%
\title{AKARI Infrared Camera Survey of the Large Magellanic Cloud. I. Point Source Catalog} 
\author{Daisuke \textsc{Kato}\altaffilmark{1,2,3},
Yoshifusa \textsc{Ita}\altaffilmark{4},
Takashi \textsc{Onaka}\altaffilmark{1},
Toshihiko \textsc{Tanab\'e}\altaffilmark{5},
Takashi \textsc{Shimonishi}\altaffilmark{1, 10},
Itsuki \textsc{Sakon}\altaffilmark{1},
Hidehiro \textsc{Kaneda}\altaffilmark{6},
Akiko \textsc{Kawamura}\altaffilmark{7},
Takehiko \textsc{Wada}\altaffilmark{2},
Fumihiko \textsc{Usui}\altaffilmark{2},
Bon-Chul \textsc{Koo}\altaffilmark{8},
Mikako \textsc{Matsuura}\altaffilmark{9},
and
Hidenori \textsc{Takahashi}\altaffilmark{5}
}
\altaffiltext{1}{Department of Astronomy, Graduate School of Science, %
The University of Tokyo, %
7-3-1, Hongo, Bunkyo-ku, Tokyo, 113-0033, Japan}
\email{kato@ir.isas.jaxa.jp; onaka@astron.s.u-tokyo.ac.jp}%
\altaffiltext{2}{Institute of Space and Astronautical Science, %
Japan Aerospace Exploration Agency, %
3-1-1, Yoshino-dai, Chuo-ku, Sagamihara, Kanagawa, 252-5210, Japan}
\altaffiltext{3}{Center for Low Carbon Society Strategy, %
Japan Science and Technology Agency, %
7, Goban-cho, Chiyoda-ku, Tokyo, 102-0076, Japan}
\altaffiltext{4}{Astronomical Institute, Tohoku University, %
6-3, Aramaki, Aoba-ku, Sendai, Miyagi, 980-8578, Japan}%
\altaffiltext{5}{Institute of Astronomy, School of Science, %
The University of Tokyo, 2-21-1 Osawa, Mitaka, Tokyo 181-0015, Japan}%
\altaffiltext{6}{Department of Astrophysics, Nagoya University, %
Chikusa-ku, Nagoya, 464-8602, Japan}%
\altaffiltext{7}{National Astronomical Observatory of Japan, %
2-21-1 Osawa, Mitaka, Tokyo, 181-8588, Japan}%
\altaffiltext{8}{Department of Physics and Astronomy, %
Seoul National University, Seoul 151-742, Korea}%
\altaffiltext{9}{Department of Physics and Astronomy, %
University College London, %
Gower Street, London WC1E 6BT, UK}
\altaffiltext{10}{Present address: Department of Earth and Planetary Sciences, 
Graduate School of Science, Kobe University, Nada Kobe 657-8501, Japan}
\begin{abstract}
We present a near- to mid-infrared point source catalog of 5 photometric bands at
3.2, 7, 11, 15 and 24 $\mu$m
for a 10 deg$^2$ area of the Large Magellanic Cloud (LMC) obtained with the
Infrared Camera (IRC) onboard the {\it AKARI} satellite.
To cover the survey area 
the observations were carried out at 3 separate seasons from 2006 May to June, 2006 October to December, and
2007 March to July.

The 10-$\sigma$ limiting magnitudes of the present survey are
17.9, 13.8, 12.4, 9.9, and 8.6 mag
at 3.2, 7, 11, 15 and 24 $\mu$m, respectively.
The photometric accuracy is estimated to be about 0.1 mag at 3.2 $\mu$m
and 0.06--0.07 mag in the other bands.
The position accuracy is 0$\farcs$3 at 3.2, 7 and 11 $\mu$m
and 1$\farcs$0 at 15 and 24 $\mu$m.
The sensitivities at 3.2, 7, and 24\,$\mu$m are roughly comparable to those of
the {\it Spitzer} SAGE LMC point source catalog, while the {\it AKARI}
catalog provides the data at 11 and 15\,$\mu$m, covering the mid-infrared
spectral range contiguously. 
Two types of catalog are provided:  a Catalog and an Archive.
The Archive contains all the detected sources, while the Catalog only includes the sources
that have a counterpart in the {\it Spitzer} SAGE point source catalog.
The Archive contains about 650,000, 140,000, 97,000, 43,000, and 52,000 sources
at 3.2, 7, 11, 15, and 24\,$\mu$m, respectively.
Based on the catalog, we discuss the luminosity functions at each band, the
color-color diagram, and the color-magnitude diagram using the 3.2, 7, and 11\,$\mu$m band data.
Stars without circumstellar envelopes, dusty C-rich and O-rich stars, young stellar objects, 
and background galaxies are located at distinct regions in the diagrams, suggesting that
the present catalog is useful for the classification of objects towards the LMC.
\end{abstract}
\section{Introduction}

The Large Magellanic Cloud (LMC) is one of the nearest galaxies.
The proximity of the LMC \citep[$\sim$50 kpc; e.g.,][]{dm-lmc}
allows us to study individual stars in detail.
The mean metallicity of the LMC is also known to be small
($\sim$1/4) compared to the solar abundance
(e.g., Luck et al. 1998),
providing us with an ideal place to study formation and evolution
of stars in low metallicity environments.

Individual stars in the LMC have been observed by
large area surveys at optical and near-infrared (NIR) wavelengths.
\citet{zaritLMC} detected about $2.4 \times 10^7$  point sources
in a $UBVI$ survey over a 64 deg$^2$ area of the LMC,
%and at NIR,
whereas \citet{kato2007} observed about $1.5 \times 10^7$ point sources
with a $JHK_s$ survey of a 40 deg$^2$ region.
At mid-infrared (MIR) wavelengths, the {\it Midcourse Space Experiment (MSX)}
carried out a survey over a 100 deg$^2$ area of the LMC \citep{egan2001, egan2003} and
detected a few thousand sources despite of the short integration time 
and low spatial resolution.
Two recent infrared satellites,
\textit{Spitzer Space Telescope} \citep{werner2004}
and {\it AKARI} \citep{murakami2007},
provided a capability of deep mid-infrared observations
with better spatial resolution.
The \textit{Spitzer} SAGE project \citep{sage} carried out
a uniform and unbiased imaging survey of a 49 deg$^2$ area of the LMC
at 4 photometric bands
from 3 to 8 $\mu$m of IRAC \citep{irac}
and 3 bands at 24, 70, and 160 $\mu$m of MIPS
\citep{rieke2004}, to which the SAGE-Spec performed follow-up spectroscopic 
observations in the MIR and far-infrared (FIR)
\citep{sage-spec, vanloon2010, woods2011}.  Recently {\it Herschel Space Observatory} also made
a survey observation over a $8\degr \times 8.5\degr$ area of
the LMC with spatial resolution higher than {\it Spitzer} in the FIR \citep{heritage}.

The {\it AKARI} Large Magellanic Cloud Survey consists of
a NIR to MIR imaging and
near-infrared spectroscopic survey toward the LMC.
The survey aims at a thorough study of the lifecycle of matters
and the star-formation history of the LMC on a galactic scale
owing to its wide spatial coverage and high sensitivities
over the wide spectral range \citep{itaLMC}.
The sensitivities are comparable to
those of the \textit{Spitzer} SAGE survey and
the {\it AKARI} survey covers the MIR spectral range 
contiguously, 
which fills the gap between the {\it Spitzer}/IRAC and MIPS bands.
In addition to the dedicated survey observations, the
entire LMC was observed as part of the all-sky survey
at 6 bands from the MIR to the FIR
\citep{ishihara2010, kawada2007}.
The MIR point source catalog at 9 and 18 $\mu$m 
and the FIR point source catalog
at 65, 90, 140 and 160 $\mu$m have been produced from
the all-sky survey observations and released to the 
public  \citep{ishihara2010, yamamura2010}.
These {\it AKARI} observations
provide 11 band data of the LMC
from the NIR to the FIR,
which give us a unique and useful database for 
the study of the LMC.

A brief description of the  {\it AKARI} LMC survey
and early results have been given in \citet{itaLMC}
based on a preliminary version of the catalog.
The catalog has been revised
by improving the data reduction procedure, which increases the
reliability and accuracy of the photometric results.
In this paper, we describe the point source catalog from
the imaging survey of the LMC from the NIR to the MIR
using the latest version of the data reduction process, which will
be released to the public.
We outline the {\it AKARI} LMC survey observations in \S \ref{sec:obs}, 
the data reduction procedures in \S \ref{sec:reduction},
and the structure of the catalog in \S \ref{sec:catalog}.
%We
The general characteristics of the catalog are presented in \S \ref{sec:quality},
and the properties of the sources in the catalog are discussed
in \S \ref{sec:analysis}. 
A summary is given in \S \ref{sec:summary}.

\section{Observations}\label{sec:obs}

The {\it AKARI} telescope has an effective aperture of 685\,mm
\citep{kaneda2007} and the whole telescope system including 
the focal-plane instruments was cooled by liquid Helium and onboard
mechanical coolers.  
The {\it AKARI} spacecraft was brought into a sun-synchronous polar orbit
at an altitude of 700\,km along the twilight zone,
similar to that of the {\it IRAS} satellite.
The LMC is located near the south ecliptic pole, where the polar orbit
has high visibility.  Taking this advantage, we observed the
main part of the LMC contiguously.

The {\it AKARI}  LMC survey was carried out with the Infrared Camera
(IRC: \citealt{onaka2007}) in the pointing observation mode 
\citep{murakami2007}.  
It has 3 independent channels:
NIR (1.8$-$5.5 $\mu$m),
MIR-S (4.6$-$13.4 $\mu$m),
and MIR-L (12.6$-$26.5 $\mu$m).
Each channel has a wide field of view (FoV) of about
10$\arcmin$ $\times$ 10$\arcmin$
and is equipped with 3 imaging filters
and 2 dispersion elements,
which can be switched during a pointing observation.
Of these, we use 5 imaging filters at
3.2 (N3), 7 (S7), 11 (S11),
15 (L15), and 24 $\mu$m (L24),
and the prism for slitless spectroscopy
in 2--5\,$\mu$m (NP) of a spectral resolution of
$\lambda / \Delta \lambda$ $\sim$ 20 \citep{ohyama2007},
where the symbol in parentheses indicates the filter designation.
The imaging bands cover the spectral range
from 5 to 26 $\mu$m contiguously and
the S11 and L15 bands fill the gap in the wavelength coverage between the IRAC
and MIPS onboard {\it Spitzer}.
The present photometric catalog is produced from the imaging data.
Part of the data taken in the slit-less spectroscopic mode are reported in
\citet{shimonishi2008}. The data reduction and the extracted spectra 
with the slit-less spectroscopic mode will be given separately \citep{shimonishi2012}.

We use the {\it AKARI} IRC AOT02 observation template
with a special option (using the prism instead of the N4 filter) prepared for the LMC survey.
The template is designed to take images with
2 filters in each channel
at 3 dithered sky positions in a pointing observation.
At each position, both short- and long-exposure data are taken.
The net integration time for a long-exposure image is typically
133, 147, and 147\,s for the NIR, MIR-S, and MIR-L channels, respectively,
and that for a short-exposure image is 14.0, 1.75, and 1.75\,s, respectively
(Table \ref{tab:survey}).
The long-to-short exposure time ratio of a pointing observation
is 9.5, 28, and 28 for the NIR, MIR-S, and MIR-L channels, respectively, which
increases the dynamic range of the observations (see \S\ref{sec:slmerge}).

The NIR and MIR-S channels share the same FoV by means of the beam splitter,
and thus N3, NP, S7, and S11 images 
are taken at the same region simultaneously.
L15 and L24 images
of the same region are taken at a different pointing observation
since their FoV is located 20$\arcmin$ away from that of the NIR/MIR-S
in the direction perpendicular to the {\it AKARI} orbit \citep{onaka2007}.
The FoV of the {\it AKARI} telescope moves
along the ecliptic meridian in the twilight zone, 
which changes its direction with the Earth's yearly round.  Therefore, 
the telescope can observe a given celestial position twice a year.
The position of the MIR-L channel relative to
the NIR and MIR-S channels is flipped in projection on the sky at a season separated by half a year
because the satellite moves in the opposite direction.
Taking account of the change in the projection direction,
we optimize the target positions to maximize the area
observed by all the three IRC channels with the overlapping region
of $1\farcm5$ for each image.

\begin{figure}[htbp]
  \begin{center}
    \includegraphics[width=.9\linewidth]{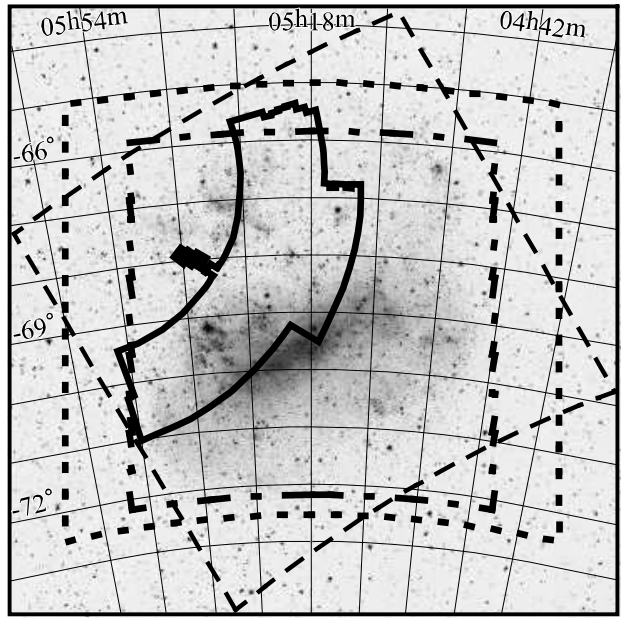}
  \end{center}
  \caption{Observed area of the {\it AKARI} IRC survey
           (solid outline) overlaid on the photographic image
           kindly provided by Motonori Kamiya.
           The dashed outline indicates
           the coverage of the {\it Spitzer} SAGE survey
           \citep[$7\degr \times 7\degr$:][]{sage},
           the dash-dotted outline shows
           the coverage of the IRSF/SIRIUS near-infrared survey
           \citep[$6.3\degr \times 6.3\degr$:][]{kato2007},
           and the dotted outline represents the coverage of
           the Magellanic clouds optical photometric survey
           \citep[$8.5\degr \times 7.5\degr$:][]{zaritLMC}.}
  \label{fig:surveyregion}
\end{figure}

Observations were carried out in 3 separate seasons,
from 2006 May 6 to June 8,
from 2006 October 2 to December 31,
and from 2007 March 24 to July 2.
Over 600 pointing observations were devoted to this project,
yielding imaging and spectroscopic maps of about a 10 deg$^2$  area
of the main part of the LMC.  Figure~\ref{fig:surveyregion} shows the
area where the {\it AKARI} survey was carried out together with the areas
of other surveys.

\section{Data reduction and compilation of
the point source catalog}\label{sec:reduction}

\subsection{Imaging data reduction}\label{sec:image}

Standard procedures for the reduction of imaging data obtained with infrared 
arrays,
dark subtraction, flat-fielding, and image co-addition, 
are performed
in the IRC data reduction toolkit version 20110304 
\citep{lorente}.
Since the basic data reduction is carried out in the standard tool kit with the default options, 
we briefly summarize the procedures below.
\begin{enumerate}

\item 
Firstly the detector non-linearity is corrected for in each exposure image.
Deviations from the ideal linear response are
estimated to be
smaller than 5\% at 12000, 20000, and 20000 ADU 
for NIR, MIR-S, and MIR-L, respectively, which roughly correspond to
the saturation levels of 0.26, 1.9, 1.8, 2.5, and 22\,Jy for the short-exposure data 
at N3, S7, S11, L15, and
L24, respectively (see also Table~\ref{tab:survey}).

\item 
The dark image is subtracted from each image.
We use the dark data taken at the beginning of the pointing observation sequence (``pre-dark'')
for the MIR-S and MIR-L long-exposure images
by averaging 3 long-exposure data (``self-dark'').
For all the short-exposure and NIR long-exposure images,
we use the ``super-dark'' images,
which are created by averaging
more than 70 ``pre-dark'' images in the LMC observations and
are provided as the default dark data in the IRC toolkit.  
To correct for the possible time variation between the ``super-dark'' and the dark image at
the time of observation, the signal level
of the ``super-dark'' image is scaled to that of the observed
image at the slit mask region before the dark subtraction,
assuming that the dark image pattern does not change with time.

\item
Signals due to high-energy ionizing particle hits (hereafter cosmic-rays)
are removed with the task COSMICRAYS in IRAF.
Remaining faint comic-ray events are removed by median-average
in step~\ref{item:combination}.

\item
Each image is divided by the flat-field image
to correct for the the pixel-to-pixel variation 
in the sensitivity over the arrays.
We employ the ``super flat-field'' images created from
images near the North Ecliptic Pole,
where the smooth zodiacal emission dominates.
We use the MIR-L flat data, in which the effect of
the artifacts by internal reflections is corrected
\citep{arimatsu2011}.

\item
The aspect ratio of each image is adjusted
to 1 to 1 by resampling the image.
After the adjustment,
the deviation from an ideal grid square is 0.2--0.3 pixel
in the NIR and MIR-S  images.
In the MIR-L images, the deviation becomes $\sim$1 pixel
at the edge of the detector.
The final pixel scale of the rescaled images is
1$\farcs$446, 2$\farcs$340 and 2$\farcs$384
for the NIR, MIR-S, and MIR-L images, respectively.

\item\label{item:combination}
Images taken at dithered positions are spatially aligned
and co-added
by taking their median to eliminate any array anomalies,
such as bad, dead, or hot pixels, and remaining cosmic-ray events.
Images are taken at least at 3 different positions in a pointing
observation in the observation template AOT02.

\end{enumerate}

\subsection{Source Detection and Photometry}
\label{sec:photometry}

We perform source detection with the IRAF%
\footnote{
IRAF is distributed by the National Optical Astronomy
Observatories, which are operated by the Association of
Universities for Research in Astronomy, Inc., under cooperative
agreement with the National Science Foundation.}/DAOFIND 
and photometry with the IRAF/DAOPHOT package
\citep{daophot} on co-added images.
We develop a procedure for the photometry
with the point spread function (PSF) fitting,
which is similar to that used by the {\it Spitzer} GLIMPSE%
\footnote{
http://www.astro.wisc.edu/glimpse/, and see the document
``Description of Point Source Photometry Steps Used by GLIMPSE'',
by B. L. Babler.} team \citep{benjamin}.
The procedure of the source detection and photometry
involves the following steps:

\begin{enumerate}

\item\label{item:daofind} DAOFIND is used to extract point-like sources
whose fluxes are more than 2\,$\sigma$ above the background.  The
extracted sources include those slightly extended compared to the PSF.

\item\label{item:apphot}
The aperture photometry is performed on the extracted
sources with an aperture radius of 10.0 pixels for N3 images,
and of 3.0 pixels for other band images using the PHOT task in IRAF.
The radius of 10.0 pix at N3 is the same as the one used for
the absolute flux calibration \citep{tanabe}.
Therefore, we do not apply aperture corrections
for the photometry at N3.
For the MIR band images, we adopt a smaller aperture
than that used in the calibration (7.5 pix) because
of the large background emission in the LMC at the MIR and
thus the aperture correction is needed
(see step~\ref{item:apcor}).
The inner radius of the sky annulus is set as the same as the aperture radius
and the width of the sky annulus is set as 5.0 pixels.

\item\label{item:pselect} 
Bright but unsaturated sources without any other sources within 7 pixels
are selected from the results of step \ref{item:apphot}
to construct the PSF.
At least 8 of such ``good'' sources are selected
in the N3, S7, and S11 images,
and 5 in the L15 and L24 images.

\item
The ``good'' sources selected in step \ref{item:pselect} are used
to construct a model PSF for each N3 image
since the PSF in the N3 images notably varies from
pointing to pointing possibly
due to jitters in the satellite pointing.
We use the PSF task in DAOPHOT to choose the best fitting function
by trying several different types of fitting functions in each image.

For the S7, S11, L15, and L24 images,
we use the common PSFs constructed in advance for each band,
instead of constructing them from each image since the effect of the jitter in
the satellite pointing is negligible compared to the size of the PSFs.
The common PSFs are constructed from about 10 ``good'' sources,
which are selected from the images
taken with the LMC survey
and the observations of 47 Tuc \citep{ita47Tuc};
sources free from diffuse emission are carefully chosen by eyes.
To construct  the common PSFs,
the ``moffat25'', ``lorentz'', ``moffat15'', and ``gauss'' models
are applied at S7, S11, L15, and L24, respectively.

\item The PSF-fitting photometry is performed for the extracted sources
in step \ref{item:apphot} using ALLSTAR in DAOPHOT in an iterative manner.
After the first ALLSTAR procedure is applied for the extracted sources, the 
sources for which photometry has been carried out are
removed from the output image of ALLSTAR.  Then 
the ALLSTAR is applied again for the remaining sources.  
We repeat the ALLSTAR procedure 3 times to have photometry for all the
extracted sources.
This is a procedure similar to the one adopted by the GLIMPSE team.

\item\label{item:apcor} The aperture correction is applied to
the PSF-fitting results.
The aperture correction factors and the PSFs of each band are assumed to be
the same among the images.
The correction factor is estimated from the median of the differences
between the results of the aperture photometry
with an aperture of 7.5 and 3.0 pixels
applied to all the sources used to construct the ``common'' PSFs
since the absolute calibration was carried out with the aperture photometry
with the 7.5 pixel radius for the MIR-S and MIR-L images \citep{tanabe}.

\end{enumerate}

We compare the results of the aperture photometry with
those of the PSF-fitting photometry.
For the aperture photometry,
only steps \ref{item:daofind}, \ref{item:apphot} and \ref{item:apcor}
above are performed.

\subsection{Photometric Calibration}\label{sec:calibration}

The resultant instrumental signals are converted into physical units
using the IRC flux conversion factors
\citep{lorente}.
Then the calibrated flux densities are converted into magnitudes
of the IRC-Vega system using the zero-magnitude flux densities
tabulated in Table \ref{tab:survey}.
The differences between the instrumental and
the calibrated magnitudes are constants.
The zero-magnitude flux densities are estimated for the short- and long-exposure images
in each band using the relative spectral response in
\citet{onaka2007}.
The IRC flux is calibrated against the 
spectral energy distribution (SED) of
$f_{\lambda} \propto \lambda^{-1}$ or $f_{\nu} \propto \nu^{-1}$.
The color correction factors for
objects with different SEDs are tabulated in \citet{lorente}.
The photometric uncertainty includesthe read-out noises, and the errors
in the aperture correction factors
and the ADU-to-Jy conversion factors.
The values of the read-out noises and
the errors of the ADU-to-Jy conversion factors
are taken from \citet{tanabe}.

\subsection{Astrometric Calibration}\label{sec:astrometry}

The attitude control system of the {\it AKARI} satellite has 
an absolute position accuracy no better than 5$\arcsec$
(the requirement was 30$\arcsec$) and thus
the astrometric calibration using other catalogs is necessary
to improve the position information of the images.
We calculate the coordinate transformation matrix that
relates the image pixel coordinates to the sky coordinates
by matching the detected point sources in the N3 images with
the Two Micron All Sky Survey Point Source Catalog
\citep[2MASS-PSC;][]{2mass},
and those in the MIR-S and MIR-L images
with the $Spitzer$ SAGE LMC point source catalog version 3.1%
\footnote{
http://ssc.spitzer.caltech.edu/spitzermission/%
observingprograms/legacy/sage/
}(SAGE-PSC; \citealt{sage}).
Sources detected in the MIR-S images are matched with the IRAC [8.0] sources,
and those detected in the MIR-L images are with the MIPS [24] sources, respectively,
where the numbers bracketed by [ ] designate
the band of the SAGE catalog; for instance,
[3.6] indicates the IRAC 3.6 $\mu$m band.
The procedure is summarized in the following.
\begin{enumerate}
\item The equatorial coordinates ($\alpha_i$, $\delta_i$)
of the 2MASS/SAGE sources in each field
are converted to ($X_i$, $Y_i$) 
in the World Coordinate System (WCS).
\item Bright sources in each long-exposure image are extracted.
\item A triangle matching between the
2MASS/SAGE sources in the WCS ($X_i$, $Y_i$) 
and the sources in the pixel coordinates ($x_j$, $y_j$) in {\it AKARI} images
is performed with the task XYXYMATCH in IRAF.
\item\label{item:calmatrix}
The transformation matrix is calculated
using the pixel coordinates ($x_i$, $y_i$)
and the equatorial coordinates ($\alpha_i$, $\delta_i$)
of the matched catalog sources with CCMAP in IRAF.
The second, third, and fourth orders of polynomials are
examined, and
then the one with the smallest residuals is adopted.

\end{enumerate}

We use at least 5 sources that match with the catalogued sources for the calculation.
For one S11, 5 L15, and 74 L24 images,
automatic matching with the SAGE catalog is unsuccessful
due to the small number of the sources in the images.
In such cases,
we match the detected point sources with the SAGE-PSC by eye.
We use only 4 sources for the manual matching and thus
the positional accuracy in these images could be worse than
those in others.
Point source matching does not always work for short-exposure images
due to the paucity of detected point sources.
For 32 S7 and 150 S11
short-exposure images, we could not obtain a sufficient number of sources and
 the same matrix calculated for the corresponding
long-exposure image is applied.
The catalog used as the position reference (2MASS-PSC or SAGE-PSC)
and the method (e.g., by eye) used for the matching 
are indicated by the flags for each source  (see \S \ref{sec:catalog}).

As described above,
the astrometric reference of the present catalog is the 2MASS-PSC and SAGE-PSC.
The equatorial coordinates of the 2MASS-PSC are based on
the International Celestial Reference System (ICRS),
and so is the SAGE-PSC via the 2MASS-PSC.
Hence,
the equatorial coordinates of the present catalog 
indirectly refer to the ICRS.

\subsection{Final Product Generation}

\subsubsection{Duplicate Source}
\label{sec:fieldmerge}

Each image has overlapping regions of $1\farcm5 \times 10\farcm0$ in its contiguous images.
Duplicate sources in the overlapping regions 
are merged together with the criterion of the spatial proximity of
$|\Delta r|$ $\leq$ 3$\farcs$0.
The flux density of the one with a better signal-to-noise ratio (S/N) is adopted and the other(s) is discarded.
The number of the duplicate sources for each source in each band
is denoted by the corresponding code (see \S \ref{sec:catalog}).
Note that in most cases the overlapping regions were observed in close pointing 
opportunities, but in some cases, they were observed in a different season.
For a variable source, therefore, the present catalog gives the flux density at a brighter phase.

\subsubsection{Merging Short-  and Long-exposure Data}
\label{sec:slmerge}

The above analysis produces 4 sets of the result of photometry
for a source; 
2 photometry methods
(aperture or PSF-fitting) and 2 exposure time data (short or long).
Taking account of the errors estimated from the 2 photometry
methods, we adopt the results of the PSF-fitting photometry for the  long-exposure
data for all the 5 band images.  For the short-exposure data, we adopt the results of the
PSF-fitting photometry for the N3 images and those of the aperture photometry
for the S7 and S11 images because reliable PSFs cannot be created
for the S7 and S11.  The consistency between the long- and
short-exposure data is confirmed by the aperture correction procedure
in step~\ref{item:apcor} in \S~\ref{sec:photometry} \citep[see also][]{tanabe}.  
The MIR-L short-exposure
data are not adopted in the production of the catalog because of
their poor S/N.

To increase the dynamic range,
we merge the photometry results of the short- and long-exposure data
with a positional tolerance of 3$\farcs$0
at the N3, S7 and S11 bands.
If a source is detected both in the short- and long-exposure images,
we adopt the one with a better S/N and discard the other.
The adopted photometry method and the exposure time for a given source are indicated
by the corresponding codes (see \S \ref{sec:catalog}).

\subsubsection{Band Merging}\label{sec:bandmerge}

The photometry results are merged to collect
5 photometric band data for a given source
by the following procedures:
(1) merge the N3 list with the S7 list
with a positional tolerance of 3$\farcs$0 to produced 
the N3-S7 list,
(2) merge the N3-S7 list with the S11 list
with a tolerance of 3$\farcs$0 to obtain the N3-S7-S11 list,
(3) merge the N3-S7-S11 list with the L15 list
with a tolerance of 5$\farcs$0 to make the N3-S7-S11-L15 list,
and (4) merge the N3-S7-S11-L15  list with the L24 list
with a tolerance of 5$\farcs$0 to obtain the final band-merged list.
The matched source is regarded as the same object and the information
is merged.  The sources not matched remain as independent sources in the list.
The number of the matched sources and
the distance from the nearest matched source
in the merging procedures (1)--(4) are given in the final catalog.
If more than one sources in the shorter wavelength list are matched in position, 
the one with the shortest distance is identified as the same source.
In such cases,  
the number of the position-matched sources is given for the source in question,
but the source(s) that is not identified as the same source is regarded as a
different source and the number of the matched sources is set as zero.
Note that the number of the position-matched sources is always given at
the longer wavelength band (Table~\ref{tab:content}).
For each of the matched sources, the position
of the shortest wavelength band is adopted as the source position.
The band used for the source position determination
is indicated by the corresponding code (see \S \ref{sec:catalog}).

\subsubsection{Artifact Identification}\label{sec:artifact}

Similarly to the {\it Spitzer} IRAC images, the
N3 images suffer from the ``mux-bleed''
(signals leaking from bright point sources, making false faint
point-like sources every 4 pixels along a row,
in which the bright sources are detected)
and the ``column pulldown'' effects
(reduction in intensity of the columns
in which bright sources are detected).
We identify and flag the suspected victims
that are located within a belt of $\pm$5 pixels wide along
the rows/columns where bright (saturated) sources are detected.
We also flag the sources located near very bright sources,
using the proximity radius of 20 pixels for N3,
and of 15 pixels for the other bands.
The flagged sources should be used with care.

The S7 and S11 images
suffer from the noticeable artifacts of bright sources
due to the internal reflections in the beam splitter
\citep{arimatsu2011}.
We flag the source located at the suspected artifact position
that can be estimated accurately from the position of the bright source.
No artifacts originating from the beamsplitter are appreciated or present at N3, L15 and L24,
and thus no flag for the artifact is given at these bands.
The format of the codes is described in \S \ref{sec:catalog}.

\subsubsection{Construction of Catalog and Archive}
\label{sec:Catalog}

We construct two kinds of catalog,
an Archive and a Catalog.
The former includes all the detected sources
and the latter contains only sources with high reliability.

The Catalog includes only sources that have counterparts
in the SAGE-PSC.
Although the band profiles of the {\it AKARI} IRC and
the {\it Spitzer} IRAC/MIPS are not exactly the same,
a source detected in the IRC band is likely to be detected
in the corresponding IRAC/MIPS band
because of their similar sensitivities (Fig.~\ref{fig:sensitivity}).
We employ the Archive of the SAGE-PSC,
which includes more sources than their Catalog,
and match the N3 sources with the [3.6] sources,
S7, S11 and L15 with [8.0],
and L24 with [24],
with a positional tolerance of 3$\farcs$0.
The Archive has more and fainter sources than the Catalog,
which may contain
potential false detection 
or sources with larger photometric uncertainties.
In total, the Catalog contains 660,286 sources and the Archive has 802,285 sources.

\section{The Catalog}\label{sec:catalog}

Table \ref{tab:catsample} shows the entries for the first 10 sources of
the Catalog.
The Catalog and the Archive will only be electronically available\footnote{
http://www.ir.isas.jaxa.jp/ASTRO-F/Observation/}.
The record for a source consists of 554 bytes.
Descriptions of each entry of the
Catalog and the Archive are summarized in Table~\ref{tab:content}.
The first 6 entries are common for all the 5 bands and the rest are the
band-specific information.  The band-specific entries are bundled together for each band
and given in the order of N3, S7, S11, L15, and L24.      For the band at which the source is
not detected, the band-specific entries are assigned as 99.999 or similar numbers.

The number of the matched sources and
the distance from the nearest matched source (Nmat and dis)
are given for the longer band.  Thus N3 does not have these entries.
They are denoted by S7Nmat and S7dis
when merging the N3 list with the S7 list.
Similarly S11Nmat and S11dis indicate those in merging the N3-S7 list
with the S11 list, L15Nmat and L15dis in merging
the N3-S7-S11 list with the L15 list,
and L24Nmat and L24dis in merging the N3-S7-S11-L15 list
with the L24 list.

The flag code consists of 7 characters, one for each flag:
``exposure'', ``photometry'', ``saturation'', ``mux bleed'',
``column pulldown'', ``artifact'', and ``multiple'' flags, respectively.
The description of the flags is summarized in Table \ref{tab:flag}.
Of these, the ``saturation'', ``mux bleed'', ``column pulldown'',
and ``artifact'' flags are set to 1 if the source has a possibility
of being affected by these phenomena (see \S \ref{sec:artifact}).
The ``exposure'' flag denotes
photometry either from the short (0) or the long (1) exposure data,
the ``photometry'' flag denotes the photometry method employed,
0 for the aperture photometry and 1 for the fit photometry
(see \S \ref{sec:slmerge}), and
the ``multiple'' flag provides
the number of nearby sources
(see \S \ref{sec:fieldmerge}).
The code is given for all the bands, but
``mux bleed'' and ``column pulldown'' are
only given at N3,
and ``artifact'' is assinged
 only at S7 and S11
(see \S\ref{sec:artifact}).

\section {General Characteristics of the Catalog}\label{sec:quality}

\subsection{Accuracy of Photometry and Astrometry}

\subsubsection{Photometric Accuracy}\label{sec:photo_accuracy}

The photometric accuracy of the present dataset is estimated from
the internal consistency of measured flux of the sources 
in the overlapping regions.
Table \ref{tab:accuracy} summarizes 
the median of the dispersions in the magnitude of the sources observed in more than one 
pointing observations.
The dispersion at N3 ($ \sim 0.09$ mag) is larger than those at S7 and S11 ($\sim$0.06--0.07).
Bright sources must include a large number of variable stars.
Thus the present dispersions are affected both by the photometric uncertainties
and the variability of the sources
and indicate upper limits of the photometric uncertainties.
Part of the large dispersion at N3 may come from the variability,
although a study of the variables based on the SAGE data does not
indicate a strong spectral dependence of the variability 
from 3 to 8\,$\mu$m \citep{vijh2009}.
It could also be attributed to the variable PSF of the N3 images due to
jitters in the {\it AKARI} satellite pointing.

\subsubsection{Astrometric Accuracy}\label{sec:astro_accuracy}

The median of the dispersions in the positions of the sources detected at
multiple pointing observations is
summarized in Table \ref{tab:accuracy}.
The dispersion at shorter wavelengths
is smaller than that at longer wavelengths except for N3
(\S \ref{sec:bandmerge}).
The dispersion at N3 is slightly
worse than those at S7 and S11.
This may originate in the distorted PSF at N3, 
which comes from aberrations of the {\it AKARI} telescope on orbit
\citep{kaneda2007}
or tracking errors in the attitude control system
since the exposure time of N3 is
by three times longer than the MIR band data.
The sources detected at multiple bands
adopt the position determined in the shortest wavelength band
despite the slightly worse dispersion at N3.

We compare the positions in the present catalog
with those in the 2MASS-PSC and the SAGE-PSC.
The systematic offsets and the standard deviations
between the corresponding positions
are listed in Table \ref{tab:astrometry}.
The deviation from the SAGE-PSC is $\sim$0$\farcs$5, which is 
slightly larger than that from the 2MASS-PSC of $\sim$0$\farcs$4.
This is attributable to the difference in the astrometric accuracy
between the SAGE-PSC of $\sim$0$\farcs$3 \citep{sage_report}
and the 2MASS-PSC of $\sim$0$\farcs$07 \citep{2mass}.
Taking account of these facts,
we estimate the uncertainty in the position of the present catalog as
$\sim$0$\farcs$4.
This value is in agreement with that of the internal consistency.
The offsets from the 2MASS-PSC and the SAGE-PSC are
consistent with each other.
The systematic offset of the present catalog from the ICRS is estimated to be
less than 0$\farcs$1.

\subsection{Detection Limit and Completeness}

\subsubsection{Limiting Magnitude}\label{sec:limmag}

\begin{center}
\begin{figure}
  \includegraphics[width=.85\linewidth]{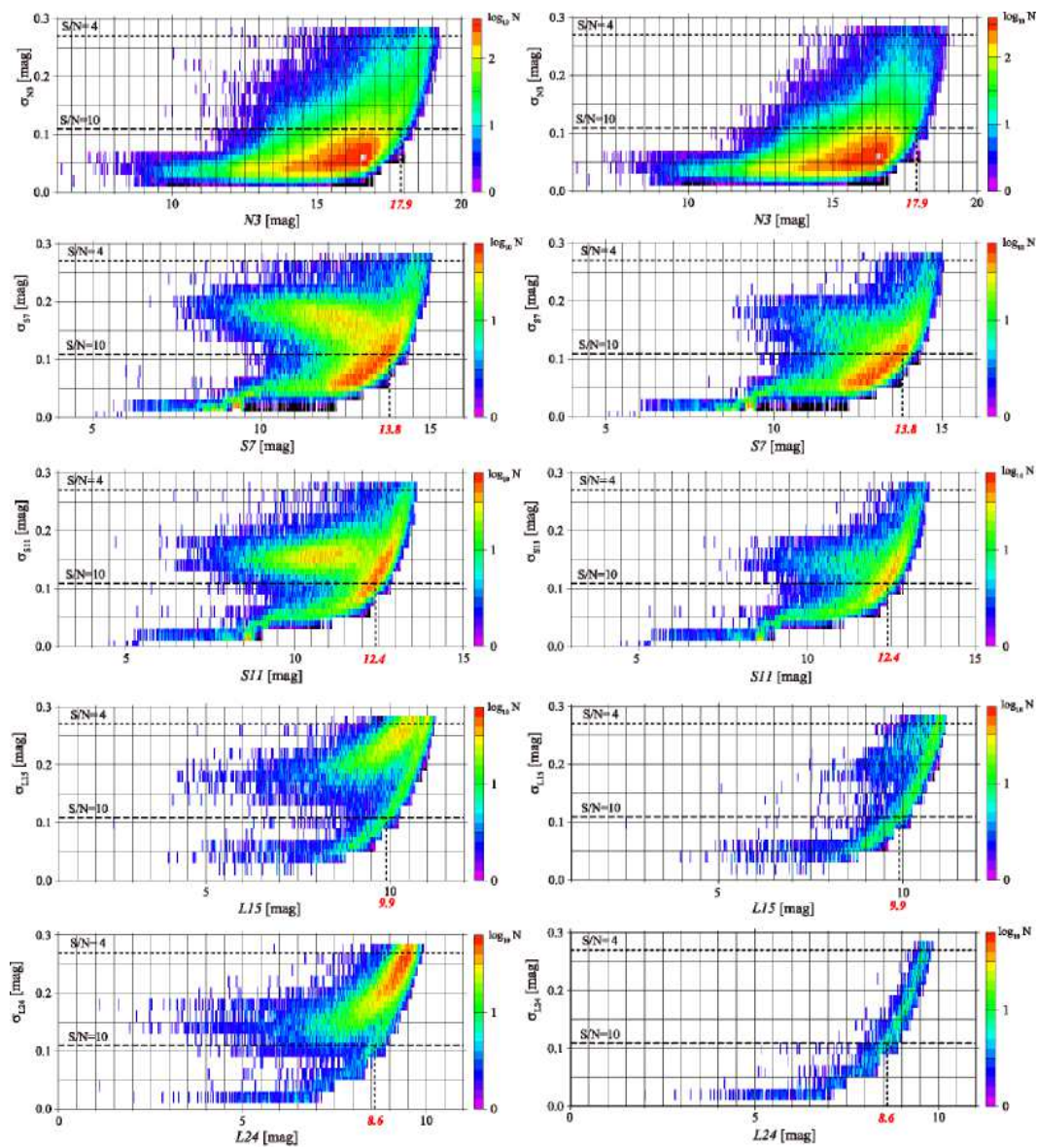}
  \caption{Photometric uncertainties as a function of magnitude
           for the Archive sources (left) and
           the Catalog sources (right)
           at N3, S7, S11, L15,
           and L24 from top to bottom.
           The horizontal dashed lines indicate
           the S/N levels of 2, 4, and 10,
           and the vertical dotted lines show
           the 10-$\sigma$ limiting magnitudes.}
  \label{fig:limmag}
\end{figure}
\end{center}

The distributions of the photometric uncertainty versus magnitude
for the sources in the Archive and the Catalog
are shown in Figure \ref{fig:limmag} for each band; 
the dashed horizontal lines indicate a S/N of 10.
The 10-$\sigma$ limiting magnitude is the faintest magnitude
at which the mode of the photometric uncertainties of the sources
in the Catalog becomes 0.11 mag.
The 10-$\sigma$ limiting magnitudes are
16.8, 13.4, 11.5, 9.9, and 8.5 mag
at N3, S7, S11, L15, and L24, respectively,
as listed in Table
\ref{tab:limit}.  
Figure \ref{fig:limmag} indicates the presence of two branches
in the sources in the Archive except for N3:
the one for which $\sigma$ goes below 0.1 mag at the brightest end
and the other in which $\sigma$ stays above 0.1 mag.
The latter branch disappears in the plot of the Catalog sources (left), suggesting
that they do not have counterparts in the SAGE catalog.
We will discuss this issue in \S \ref{sec:cmp_sage}.

\begin{figure}
  \includegraphics[width=.95\linewidth]{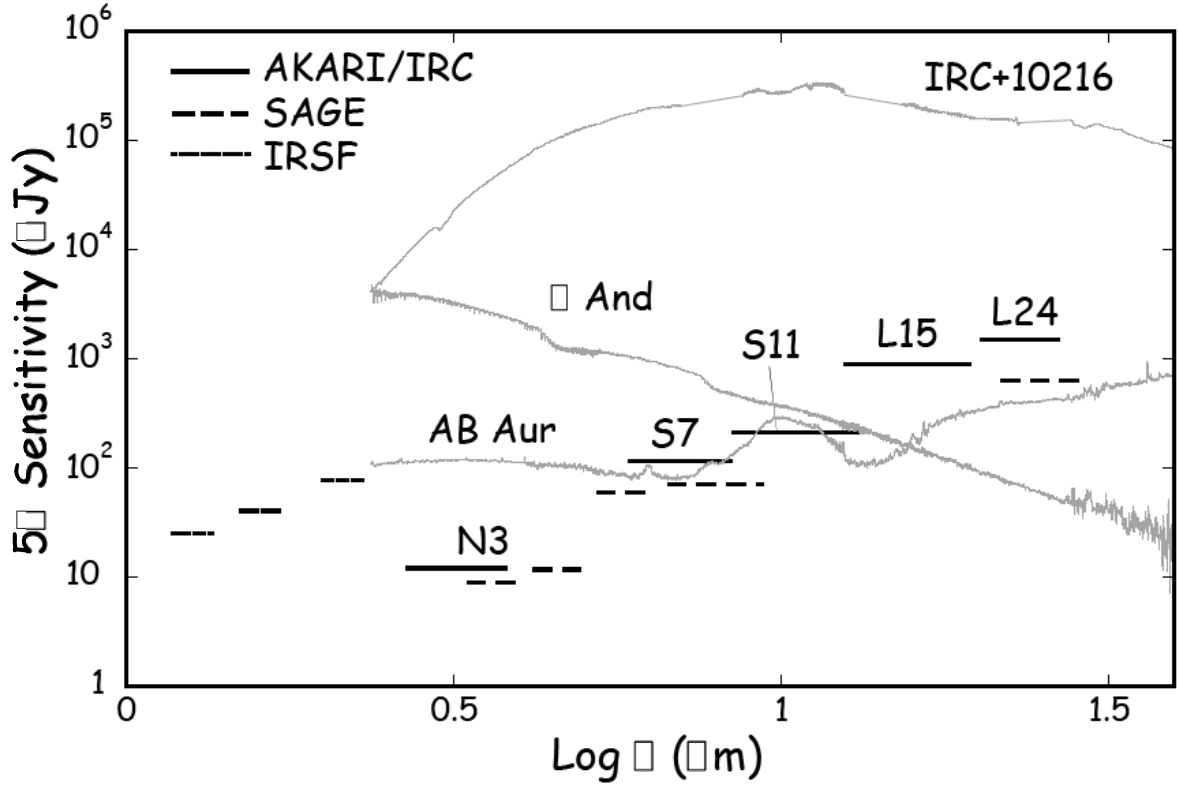}
  \caption{Graphic representation of the 5-$\sigma$ detection
           limits of the {\it AKARI} LMC survey (solid lines).
%           and the All-Sky Survey (magenta).
           For a comparison, the 5-$\sigma$ detection limits of
           the {\it Spitzer} SAGE survey and the IRSF/SIRIUS survey are
           shown by the dashed and dotted lines, respectively.  The length
           of the bars indicates the FWHM of the bandpass.
           ISO SWS spectra \citep{iso}
           of 3 galactic stars are also plotted after scaling their fluxes at the distance
           of the LMC
           as examples of a Herbig Ae/Be star (AB Aur),
           a red giant with luminosities below the TRGB ($\beta$\,And),
           and a dusty red giant (IRC +10216).}
  \label{fig:sensitivity}
\end{figure}

Figure \ref{fig:sensitivity} is a diagrammatic representation of
the 5-$\sigma$ limiting magnitudes of the present survey,
the {\it Spitzer} SAGE survey, and the IRSF/SIRIUS near-infrared survey.
The 5-$\sigma$ limiting magnitudes of the present survey are
calculated from the 10-$\sigma$ limiting magnitudes
given in Table~\ref{tab:limit}.
We estimate the 10-$\sigma$ limiting magnitudes for the SAGE survey
by the same procedure as described above,
then calculate the 5-$\sigma$ limiting magnitudes.
Figure \ref{fig:sensitivity} indicates that the limiting magnitudes
of the present survey are slightly worse than those of the SAGE survey,
while the S11 and L15 bands fill the gap of the SAGE data with comparable
sensitivities.
Examples of the {\it ISO} SWS spectra of stars of known distance
(the distances are taken from \citet{crosas97} for
IRC +10216 and \citet{perryman97}  for others) are also plotted
after scaling their flux densities
at the distance of the LMC.
Here, the distance modulus of the LMC is assumed to be 18.5 mag.
The figure indicates that all red giants
above the tip of the first red giant branch (TRGB) 
and some fraction of Herbig Ae/Be stars
in the LMC can be detected at shorter than 11\,$\mu$m in the present survey.

\subsubsection{Completeness Limit}\label{sec:cmplim}

The completeness limit is estimated by
adding artificial sources of various magnitudes
and measuring the detection probability of the sources.
Artificial sources with the shape of the PSF-model
are added by an IDL software.
90\% probability is adopted as the completeness limit.
To estimate the completeness limits
with different sky-background levels,
the following three cases are investigated in each band:
(1) ``high'': images with the highest sky-background level,
(2) ``medium'': images with the median of
the sky-background level of all images, and
(3) ``blank'': images with the lowest sky-background level.
For each level, 5 images without large masked areas or artifacts
are selected and artificial sources are embedded.

The averages and standard deviations of the 90\% completeness limits
estimated from 5 images at each level
are listed in Table~ \ref{tab:limit}.
In all bands, the 90\% completeness limits with the ``medium'' level are
comparable to those with the ``blank'' level
and those with the ``high'' level are by $\sim$3 mag
shallower than those with the other two.
The ``high''  sky-background level is higher by
3 standard deviations than the median of
the sky-background levels of all images in each band.
The fraction of the images with the ``high'' sky-background level
is only 1--2\%.
More than 90\% of the images have the sky-background level
within two standard deviations from the median of
the sky-background level of all the images at each band.
Therefore, the typical completeness limits are
comparable to those with the ``medium'' level.
The 90\% completeness limits with the ``medium'' level
are 14.6, 13.4, 12.6, 10.7, and 9.3 mag
at N3, S7, S11, L15, and L24, respectively.
They are roughly comparable with
the 10-$\sigma$ limiting magnitudes
except at N3, where it is by $\sim$2 mag shallower than
the 10-$\sigma$ limiting magnitudes.
This can be attributed to confusion due to the
relatively large number of sources at N3.

\subsection{Comparison with \textit{Spitzer} SAGE Point Source Catalog}
\label{sec:cmp_sage}

As described in \S \ref{sec:Catalog},
we construct two kinds of catalog:
the Archive, which includes all the detected sources, and
the Catalog, which includes sources that have a counterpart(s)
in the closest band of the SAGE-PSC.
Therefore, the ratio between the number of the sources in
the Catalog and that in the Archive in a band is the fraction
of the {\it AKARI} sources that have a counterpart(s) in the SAGE-PSC.
The ratios are 85.0, 56.8, 43.4, 25.1, and 4.8\%
at N3, S7, S11, L15, and L24, respectively.
Since a large fraction of the area of the {\it AKARI} survey is covered by the SAGE
survey and the sensitivities are comparable (see \ref{sec:limmag}),
the ratios are unexpectedly small except for N3.

\begin{center}
\begin{figure}
  \includegraphics[width=.85\linewidth]{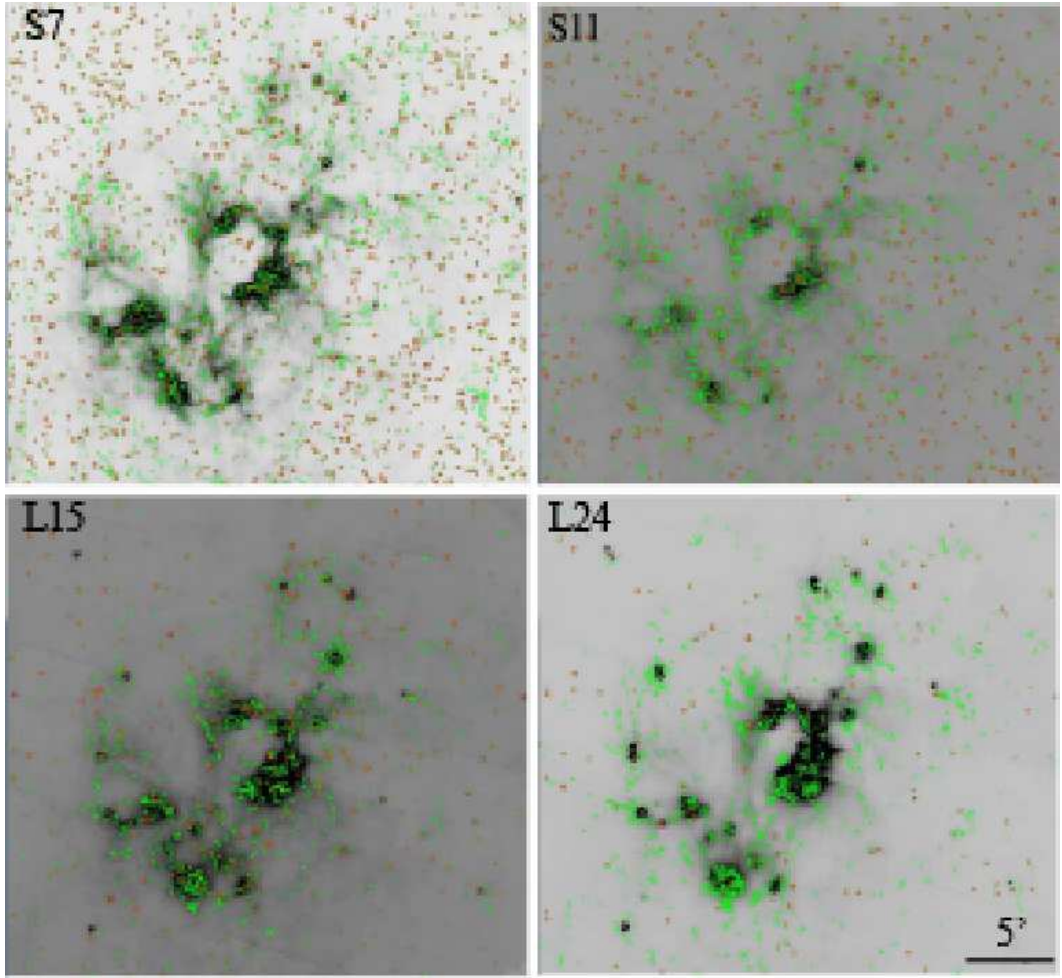}
  \caption{Spatial distributions of Archive (green) and
          Catalog sources (red) on the S7 (top left),
           S11 (top right), L15 (bottom left), and
           L24 (bottom right) images around bright nebulae.
           The image size is $30\arcmin \times 30\arcmin$ and the
           central position of the image is (5h22m08s, $-67$d56m12s).}
  \label{fig:S7plot}
\end{figure}
\end{center}

\begin{center}
\begin{figure}
  \includegraphics[width=.85\linewidth]{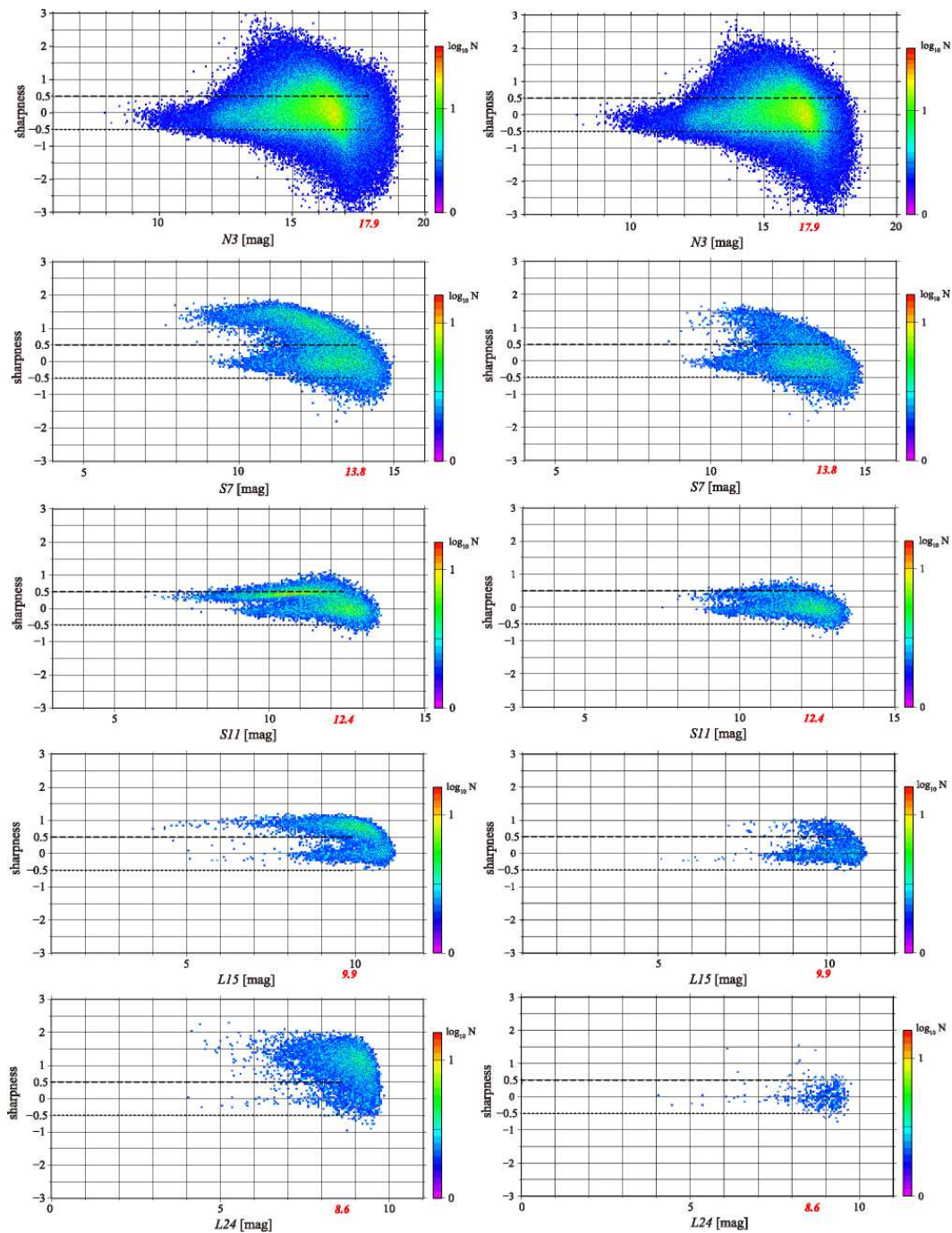}
  \caption{Sharpness vs. magnitude.  The left figure shows
           the Archive sources and the right indicates
           the Catalog sources
           at N3, S7, S11, L15 and L24 from top to bottom.
           The numbers in red indicate the 10-$\sigma$ limiting magnitudes.}
  \label{fig:sharp}
\end{figure}
\end{center}

The spatial distributions of the Archive and the Catalog sources
in S7, S11, L15, and L24 images around bright nebulosity are
shown in Figure \ref{fig:S7plot} to investigate the difference in more detail.
It indicates that the
sources only in the Archive
are concentrated along bright nebulosity. 
Figure \ref{fig:sharp} shows the relationship between the sharpness and
the magnitude for the Archive and Catalog sources
at N3, S7, S11, L15, and L24.
The sharpness parameter indicates the size of the source compared to the PSF.
A positive (negative) sharpness value indicates that the source is more (less)
extended than the PSF.
The Archive sources at S7, S11, L15, and L24 show distributions
around unity,
suggesting that they are more extended than
point-like sources.
These distributions are not seen in the Catalog sources.

Table \ref{tab:match} summarizes the number of the sources with a counterpart
in the neighboring shorter wavelength band as well as the number of all the sources
in the Archive and the Catalog.
The S7 sources are matched with the N3 sources, S11 with S7,
L15 with S11, and L24 with L15.
A significant fraction of the Archive sources have a counterpart
in the neighboring shortest wavelength band for S7, S11, L15, and L24.
These facts suggest that the Archive sources are real, 
but that they are slightly extended for the beam of {\it Spitzer} and thus are not
included in the SAGE-PSC as a true point source.
Figure~\ref{fig:limmag} indicates that the Archive sources excluded
in the SAGE catalog have larger flux uncertainties than those in the Catalog except for N3.
The larger uncertainties may be attributable to the nebulosity with which the sources are
associated as suggested by the larger sharpness. 
The associated nebulosity is appreciably seen in the MIR and increases the background fluctuation.
The imperfect match with the PSF for slightly extended sources could also increase the
uncertainties.

\begin{center}
\begin{figure}
  \includegraphics[width=.45\linewidth]{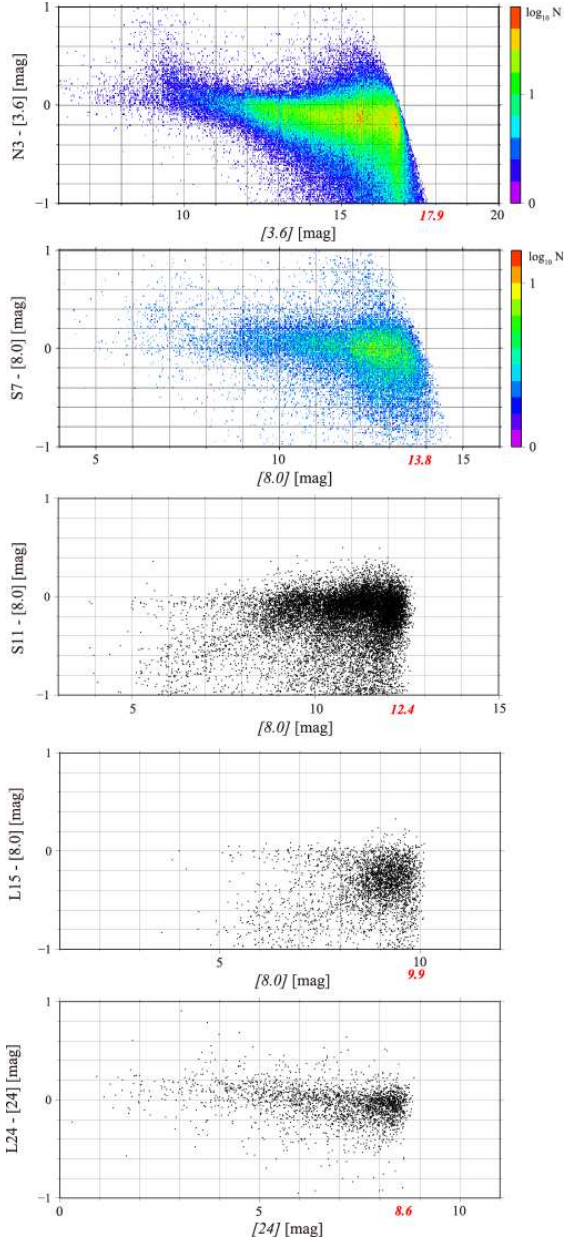}
  \caption{Comparison of the magnitudes of the {\it AKARI} catalog and SAGE-PSC.
           Residual magnitudes of the sources matched
           between the AKARI catalog and SAGE-PSC are shown
           as a function of magnitude
           at N3 (with SAGE [3.6]),
           S7, S11, L15 (with [8.0]),
           and L24 (with [24]) from top to bottom.
           The numbers in red
           indicate the 10-$\sigma$ limiting magnitudes
           of the {\it AKARI} catalog.}
  \label{fig:matsage}
\end{figure}
\end{center}

Figure \ref{fig:matsage} shows the distributions of the difference
between the N3, S7, and L24 magnitudes
and the [3.6], [8.0], and [24] in the SAGE-PSC for the Catalog sources.
Whereas systematic differences are present, which can be attributed to the difference
in the relative spectral responses, all the bands show fairly linear correlations,
confirming the consistency in the photometric results between the two datasets.

\section{Properties of the Sources in the Catalog}\label{sec:analysis}

In this section, we discuss the following properties of the sources in the
present catalog:
(1) luminosity functions (LFs), (2) a color-color diagram (CCD),
and (3) a color-magnitude diagram (CMD).

\subsection{Luminosity Functions}\label{sec:lf}

\begin{figure}
  \includegraphics[width=.43\linewidth]{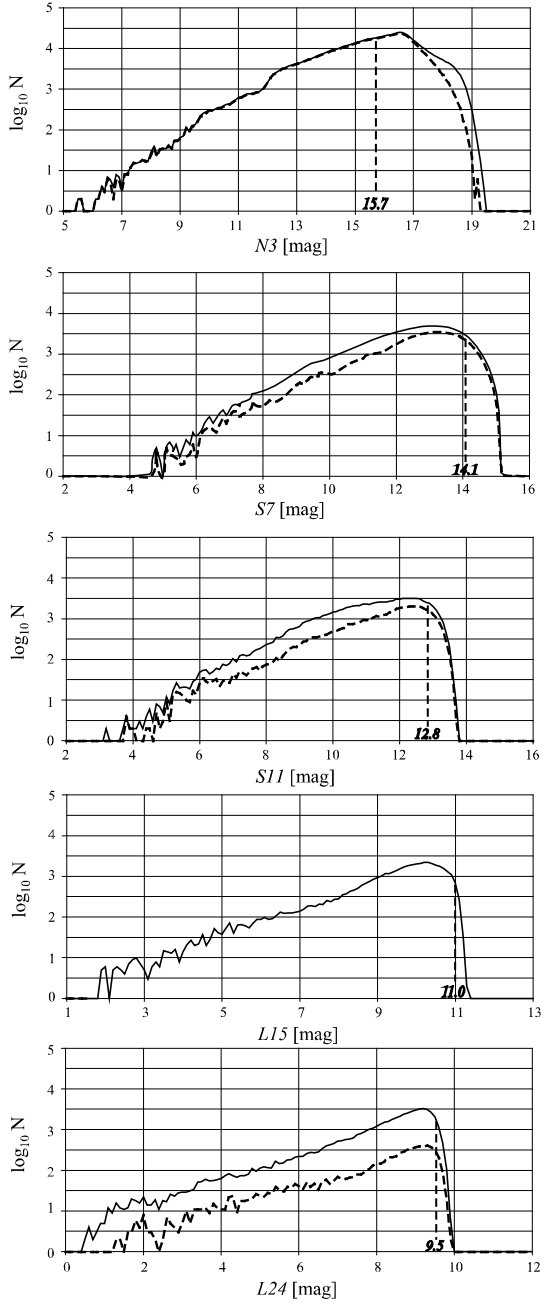}
  \caption{Luminosity functions.  From tom to bottom those at
           N3, S7, S11, L15,
           and L24 are shown in 0.1 mag bins.
           The solid and dashed lines
           show
           the Archive and Catalog sources, respectively.
           The numbers in italic and the vertical dashed lines
           indicate the 90\% completeness limits in ``blank'' regions.
}
  \label{fig:lf}
\end{figure}

\begin{figure}
  \includegraphics[width=.5\linewidth]{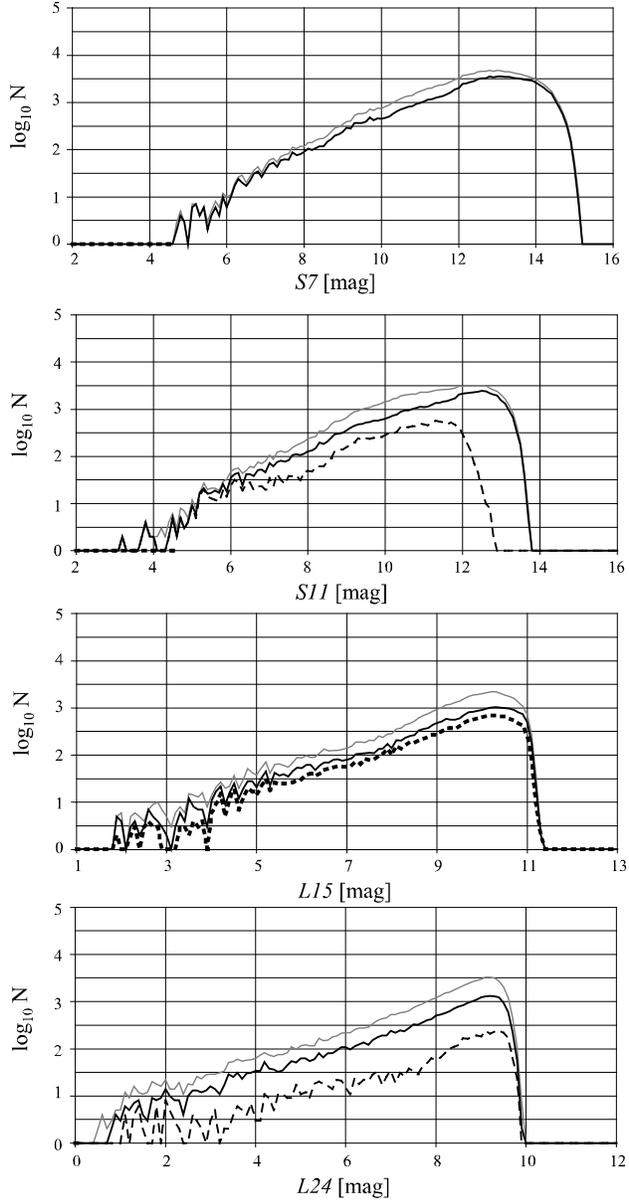}
  \caption{Luminosity functions of the Archive sources.  Those at
           S7, S11, L15,
           and L24 are shown from top to bottom.
           The gray lines indicate
           all Archive sources, the black solid lines
           those with a counterpart in N3 sources,
           and the dashed and dotted lines 
           those with a counterpart of N3 
           brighter and fainter than 12 mag, respectively.
}
  \label{fig:lfdivN3}
\end{figure}

The LFs for the 5 bands are shown
in Figure \ref{fig:lf}.
For the magnitudes brighter than 2--4 mag,
all of the LFs start to fall rapidly due to the saturation.
In the N3 LF, there is a stepwise increase around 12 mag,
which corresponds to the flux of TRGB \citep{itaLMC}.
To investigate the populations of the sources that contribute to the LFs at L15 and L24,
we also plot the Archive sources that have N3 counterparts brighter and fainter than
12 mag separately by the dotted and dashed lines in Figure~\ref{fig:lfdivN3}. 
In the LFs of S7 and S11, there are no apparent differences depending on the
brightness of the N3 counterparts brighter than 11.5 mag, but the sources
fainter than 12 mag at N3 become dominant for those fainter than 11.5 mag.
In the LFs of L15 and L24, the sources fainter than 12 mag at N3 are dominant
irrespective of the magnitude.
At the present sensitivities, TRGB without dust shells cannot be
detected at L15 and L24 (Figure~\ref{fig:sensitivity}).  Therefore, those that
contribute to the LFs of L15 and L24 are likely to be dusty AGB stars whose
N3 is significantly attenuated.
Other types of objects, such as background galaxies, are also supposed to be present.

\subsection{Color-Color Diagram}\label{sec:2cd}

\citet{itaLMC} discuss the characteristics of the CCD of  $[N3]-[S11]$ vs. $[S11]-[L15]$ using the
preliminary  {\it AKARI} LMC catalog, where the brackets $[\,]$ indicate the magnitude at the band
in question.  Based on the same catalog,
\citet{siudek2012} present the distributions of about 2000 sources that are detected at all the IRC 5 bands
and have information of the  object type on CCDs and CMDs of various combinations.
They also discuss the classification of objects using the Support Vector Machine algorithm.
These investigations demonstrate the importance of the bands S11 and L15 in the
classification, which are unique to {\it AKARI}.  \citet{itaLMC} suggest that
even the N3 data contain different information from the IRAC 3.6\,$\mu$m band.
The short cutoff wavelength of the N3 band is 2.7\,$\mu$m \citep{onaka2007}, much shorter than
that of the IRAC 3.6\,$\mu$m band \citep{irac} of 3.185\,$\mu$m.
The N3 band is more sensitive to the absorption at 3\,$\mu$m either of water ice seen 
in young stellar objects (YSOs)
or of C$_2$H$_2$ and HCN in carbon stars.
Although it contains interesting information, L15 may not have a sufficient depth to probe low-mass YSOs 
(Figure~\ref{fig:sensitivity} and see below).  
Therefore we investigate the CCD of $[N3]-[S7]$ vs. $[S7]-[S11]$, a combination of the most sensitive
3 bands, here in some detail.

\begin{figure}
  \includegraphics[width=.7\linewidth]{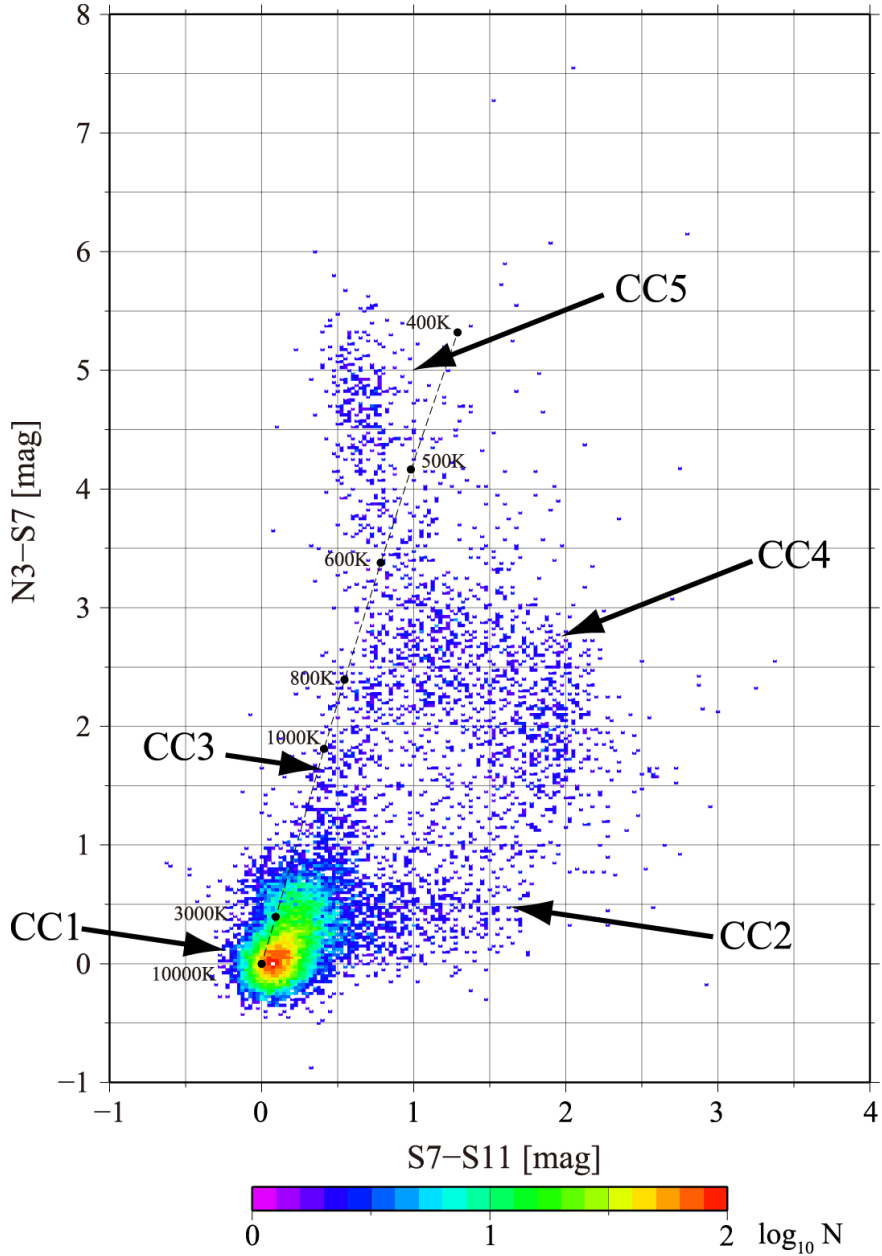}
  \caption{Color-color diagram of  $[N3]-[S7]$ vs. $[S7]-[S11]$.
           The Archive sources with S/Ns $\ge$ 10 are shown.
           Each color is binned by 0.025 mag and
           the number density levels are shown in a logarithmic scale.
           The dashed line is a locus of the colors for
           blackbodies from temperatures of 400 to 10000K.
           Noticeable features are labeled as CC1 to CC5 (see text).}
  \label{fig:2cdall}
\end{figure}

Figure~\ref{fig:2cdall} shows
the $[S7]-[S11]$ vs. $[N3]-[S7]$ CCD
for the Archive sources with S/Ns $\ge 10$ at N3, S7, and S11.
The sample contains 16446 sources.
In the CCD,
there are 5 noticeable features, which are labeled
as CC1--CC5.
Group CC1 forms a clump around $([S7]-[S11],\, [N3]-[S7]) \sim (0, 0)$.
Groups CC2 and CC3 constitute sequences extending from CC1 to redder regions;
CC2 is a horizontally stretching feature to $([S7]-[S11], \,[N3]-[S7]) \sim (3, 1)$, whereas
CC3 is a vertical sequence to $\sim (1, 2.5)$.
CC4 forms a diffuse concentration
around $([S7]-[S11], \,[N3]-[S7]) \sim (1.5, 2.5)$ and
CC5 is a sequence vertically extending from CC3 further to $([S7]-[S11], \,[N3]-[S7]) \sim (0.7, 5.5)$.

\begin{figure}
  \includegraphics[width=.45\linewidth]{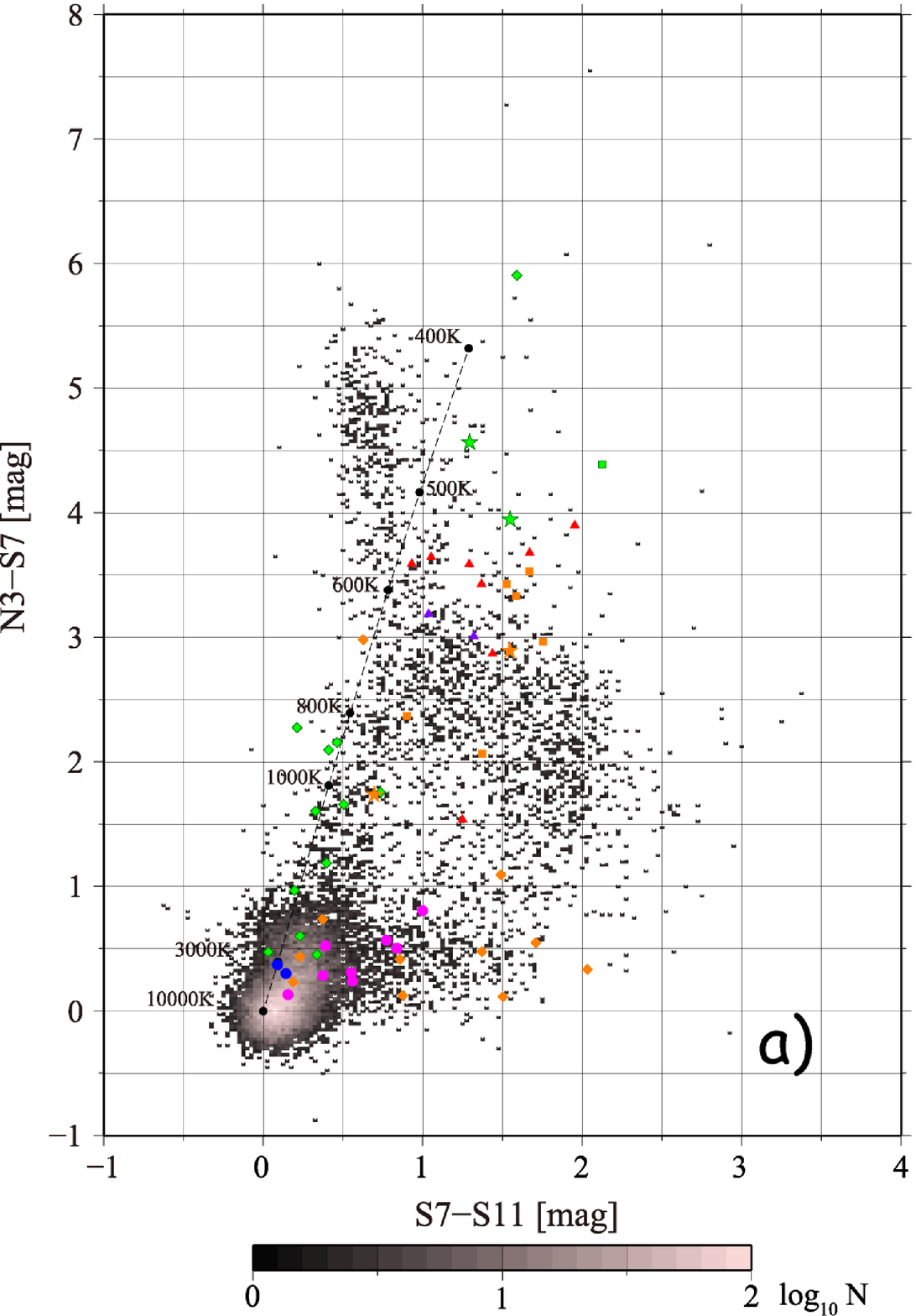}
 \includegraphics[width=.45\linewidth]{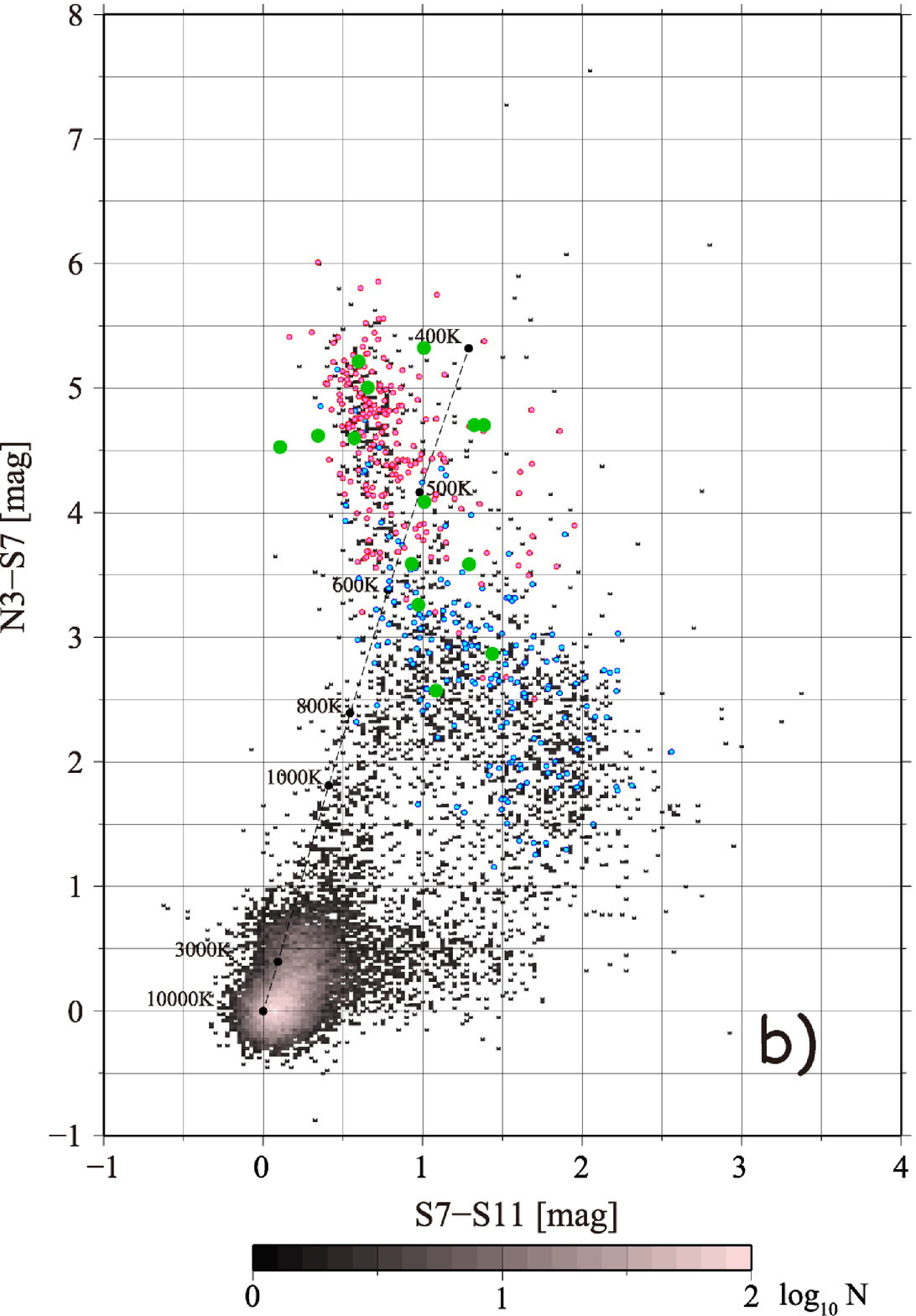}
  \caption{Color-color diagram of  $[N3]-[S7]$ vs. $[S7]-[S11]$.
           The Archive sources with S/Ns $\ge$ 10 are plotted in
           a gray-scale together 
           with  the sources classified in
           \citet{woods2011} in (a)
           and with the YSO and background galaxy candidates
           selected by \citet{yso_gruendl} in (b).
           In (a),
           the blue and purple circles indicate naked stars and red supergiants, 
           the orange and green diamonds represent O-rich and C-rich AGB stars, 
           the orange and green squares show O-rich and C-rich post-AGB stars,
           the orange and green stars indicate O-rich and C-rich planetary nebulae,
           and the purple and red triangles show \ion{H}{2} regions and
           YSOs, respectively.
           In (b),
           the blue and red circles show
           background galaxy and YSO candidates, respectively.
           The green circles indicate YSOs with the water ice absorption at 3\,$\mu$m
           \citep{shimonishi2010, shimonishiPhD}.
}
  \label{fig:2cdyso}
\end{figure}

\citet{woods2011} classify 197 point sources in the LMC based on SAGE-spec observations
with the Infrared Spectrograph (IRS) onboard {\it Spitzer} \citep{houck2004}.
With a positional tolerance of 10$\farcs$0, 67 sources in \citet{woods2011} are
found to match with the Archive sources.  
Figure~\ref{fig:2cdyso}a highlights these matched sources by the large symbols.  
Two naked stars matched (blue circles) are located at CC1 and red supergiants (purple circles)
are distributed over CC1 and CC2.  O-rich AGB stars (orange diamonds) are located at CC2,
while C-rich AGB stars (green diamonds) are located at CC3 and CC5.  O-rich post-AGB 
stars (orange squares) and O-rich planetary nebulae (orange stars) are distributed in the CC4 region.  
Only one C-rich post-AGB star and one C-rich planetary nebulae are matched with
the Archive sources and they are located at CC5 as indicated by the green square and star, 
respectively.  Two \ion{H}{2} regions are matched (purple triangles), which are seen in the CC4--CC5 regions.
No background galaxies in their sample are matched with the Archive.  YSOs (red triangles)
are located at CC5.  It should be noted that these YSOs are most likely massive ones because
of the detection limit of the IRS \citep{woods2011}.
Sources in CC4 and CC5 also include post-AGB stars and planetary nebulae.

In Figure~\ref{fig:2cdyso}b,
background galaxy and YSO candidates selected by \cite{yso_gruendl}
are plotted on the same CCD together with the Archive sources.
\cite{yso_gruendl} select the candidates
based on the analysis
using the images and photometry of the \textit{Spitzer} SAGE survey
as well as other surveys from optical to NIR wavelengths.
With a positional tolerance of 3$\farcs$0,
379 YSO and 239 background galaxy candidates
are matched with the Archive sources.
Figure~\ref{fig:2cdyso}b shows that CC5 selectively overlaps
with YSO candidates, while
background galaxies and other YSOs are distributed around the CC4 region.
It also indicates
a clear separation between YSO and background galaxy candidates
around $[N3]-[S7] \sim 3.5$, which
is roughly consistent with
the boundary between CC4 and CC5. 
The color $[N3]-[S7]$ becomes very red for YSOs because of the water ice absorption
at 3\,$\mu$m \citep{shimonishi2008}, whereas $[S7]-[S11]$ stays blue due to the
silicate absorption \citep{woods2011}.  To confirm the effect of the water ice absorption,
the YSOs that show the absorption \citep{shimonishi2010, shimonishiPhD} are also plotted
by the green circles in Fig.~\ref{fig:2cdyso}b.  11 of the 14 YSOs are located in the CC5 region, while
the rest 3 are in the CC4 region, supporting the above interpretation.
The S7 band efficiently probes the unidentified
infrared (UIR) bands at 6.2 and 7.7\,$\mu$m \citep{onaka2009}, which also accounts partly for
the redness of $[N3]-[S7]$.  Therefore, the color $[N3]-[S7]$ 
is efficient to distinguish YSOs from dusty AGB stars.  Further observations of the objects
in the CC5 region would be interesting to study their nature and investigate the applicability of
this color diagnosis.
Background galaxies can
have very red  $[S7]-[S11]$ color, but their $[N3]-[S7]$ color may not be very red
because stellar components contribute to N3. 
Post-AGB stars or planetary nebulae
are also suggested to be distributed in the CC4 region \citep{siudek2012}.

In summary, the features CC1 to CC5
in the $[S7]-[S11]$ vs. $[N3]-[S7]$ CCD
correspond to
the following objects:\\
CC1: naked stars;\\
CC2: dusty O-rich stars;\\
CC3: dusty C-rich stars;\\
CC4: YSOs and background
         galaxy candidates, and post-AGB and planetary nebulae;\\
CC5: YSO candidates with water ice absorption\\
These classifications are also confirmed by the investigation of known objects towards the LMC
\citep{siudek2012}.

Figure \ref{fig:2cdall} shows the distributions of the sources  similar to those seen
in the $[S11]-[L15]$ vs. $[N3]-[S11]$ CCD
based on the previous version of the {\it AKARI} LMC catalog \citep{itaLMC}
except for CC5, which is not seen in their diagram.
Their classifications of each feature are in agreement with the present
results.  The absence of CC5 in the $[S11]-[L15]$ vs. $[N3]-[S11]$ CCD
may suggest that the present sensitivity of $L15$ is  not sufficient to detect
most YSOs in the LMC.

\subsection{Color-Magnitude Diagram}\label{sec:cmd}

\begin{figure}
  \includegraphics[width=.7\linewidth]{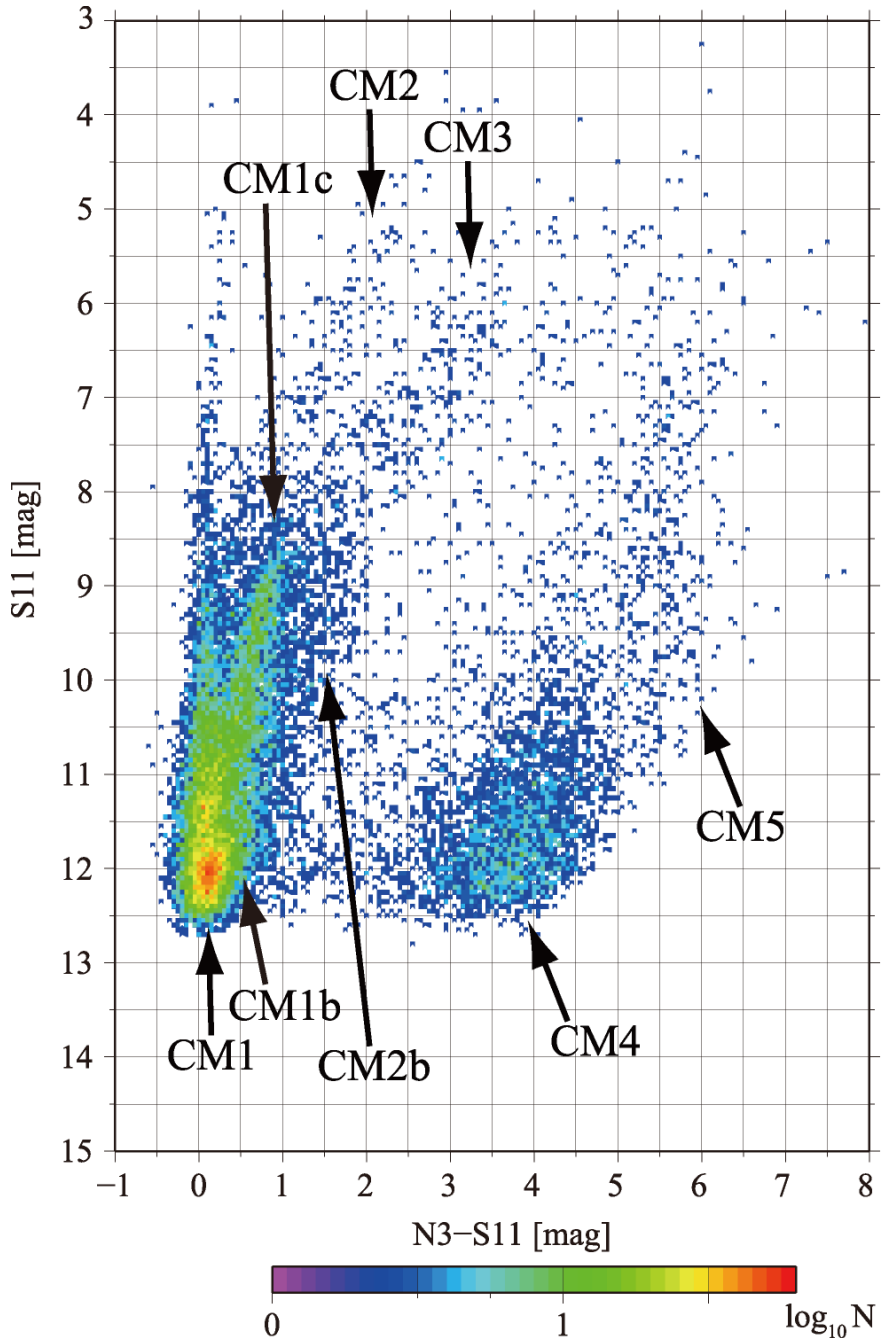}
  \caption{Color-magnitude diagram of $[S11]$ vs. $[N3]-[S11]$ 
           The Archive sources with S/Ns $\ge$ 10 are plotted.
           Each data point is binned by 0.05 mag and
           the number density levels are shown in a logarithmic scale.
           Noticeable features are labels as CM1 to CM5 (see text).
}
  \label{fig:cmdall}
\end{figure}

Classification based on the CMDs has been extensively discussed for the
SAGE-SPC data \citep[e.g.,][]{blum2006, yso_whitney}.
\citet{itaLMC} present CMDs of several combinations of the imaging bands
based on the previous version of the {\it AKARI} catalog.
Here we only briefly discuss a CMD, which shows a well-separated distribution
of various objects to confirm the previous results.
Figure \ref{fig:cmdall} shows
the $[N3]-[S11]$ vs.\ $[S11]$ CMD
for the Archive sources with S/Ns $\ge$ 10 at N3, S7, and S11.
Comparison 
with the 5 components CC1--CC5
in the CCD indicates:\\
(1) CC1 sources form a vertically stretching feature along $[N3]-[S11] \sim 0$ (CM1)
and two separate branches in $ 0 < [N3]-[S11] < 1.5$ (CM1b and CM1c);\\
(2) CC2 sources form a finger-like feature (CM2) and
a clump around $([N3]-[S11], \,[S11]) \sim (1.5, \,9.5)$ (CM2b);\\
(3) CC3 sources form a finger-like structure (CM3);\\
(4) CC4 sources form a clump (CM4);\\
(5) CC5 sources are distributed to the redder region (CM5).

The sequence CM1b corresponds to a bristle-shape feature, which
\citet{itaLMC} attribute to the broad emission feature of aluminum oxide grains
around 11.4\,$\mu$m \citep[e.g.,][]{onaka89}.  

Comparison with their CMD suggests that the distributions of dusty C-rich AGB stars are
located at the position of CM3
and those of optical carbon stars correspond to the sequence CM1c
($([N3-S11], \,S11) \sim (0.5, \, 10.5)$ to (1.0, 8.5)).
Other features are also basically seen 
in the CMD
of \citet{itaLMC}, confirming the previous results by the improved catalog.

We find a new feature CM2b in Figure~\ref{fig:cmdall}.
The sources of CM2b are found to correspond to CC2 in the CCD (Figure~\ref{fig:2cdall}), which
we assign to
dusty O-rich stars (see Figure~\ref{fig:2cdyso}a).
According to stellar evolution models
\citep[e.g.,][]{3du2, 3du3,3du4},
the third dredge-up enhances the abundance of carbon relative to oxygen for stars with $\ge 2 M_{\odot}$,
which leads to the formation of C-rich stars \citep{3du1}.  
For stars with $\ge 4 M_{\odot}$, the hot bottom burning 
decreases the abundance of carbon,
which then prevents the formation of C-rich stars \citep{hbb1}.
These suggest that AGB stars with an intermediate-mass ($\sim$2--4 M$_{\odot}$) 
have a high probability of evolving to C-rich stars, and
low-mass ($\le 2 M_{\odot}$) and high-mass
($\sim$4--8 $M_{\odot}$) AGB stars
are likely to evolve to O-rich stars.
Thus theoretical predictions suggest that
CM2 and CM2b in Figure~\ref{fig:cmdall}
correspond to high-mass and low-mass O-rich stars, respectively.
Note that both mass limits 
depend on the metallicity \citep{hbb2} and thus the location of CM2b may differ in
other galaxies.  

\begin{figure}
 \includegraphics[width=.45\linewidth]{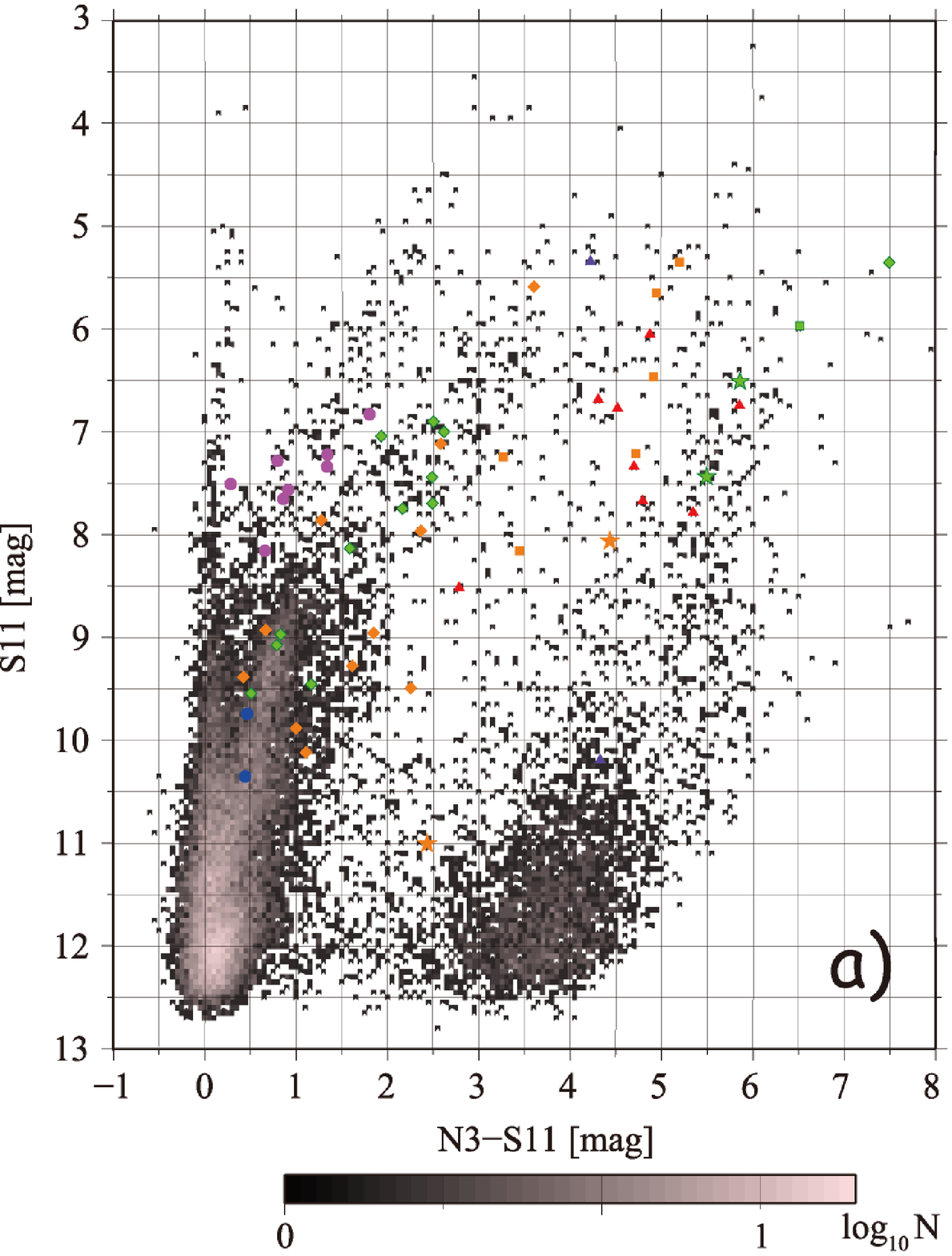}
 \includegraphics[width=.45\linewidth]{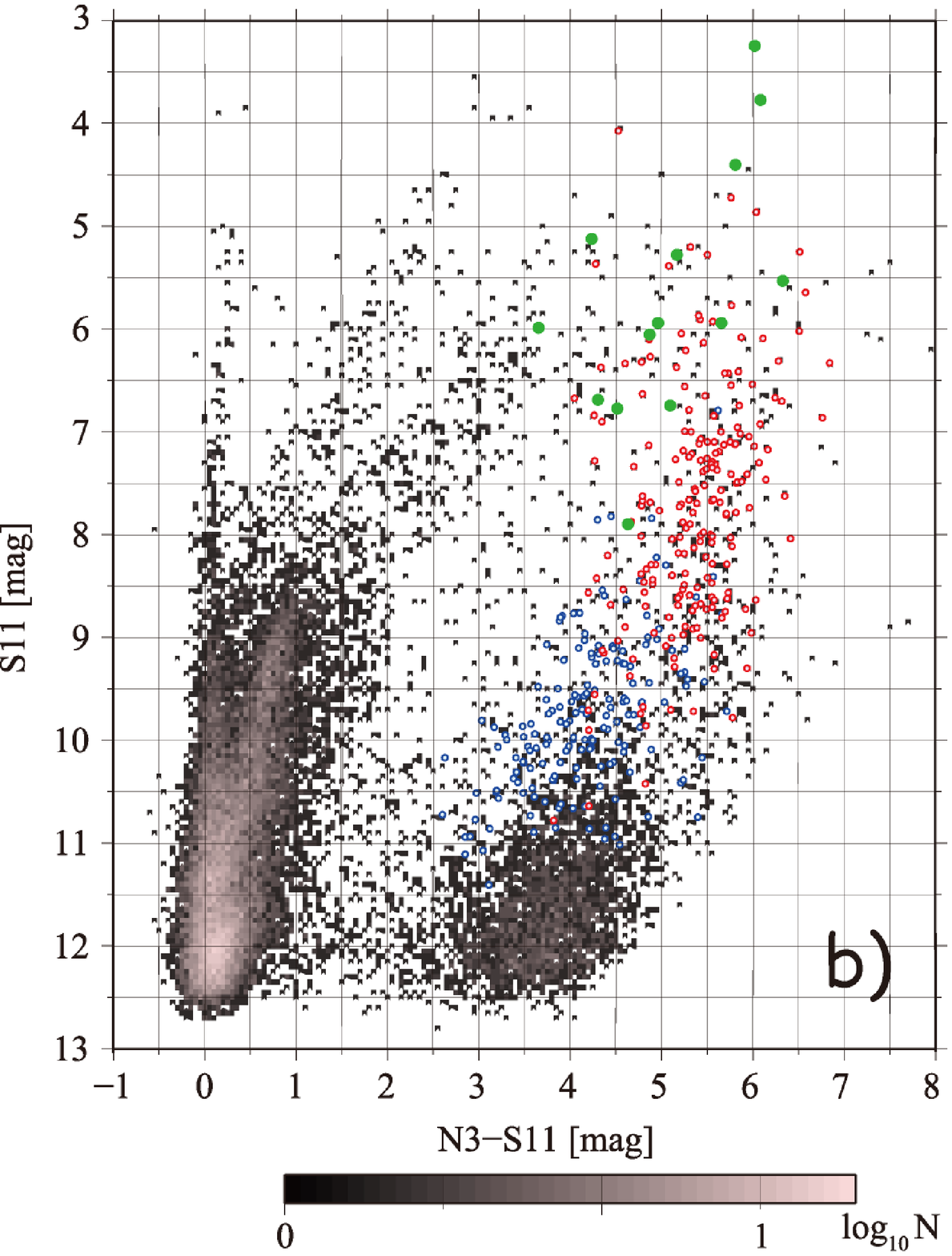}
  \caption{Color-magnitude diagram of $[S11]$ vs. $[N3]-[S11]$.
           The Archive sources with S/Ns $\ge$ 10 are plotted in a
           gray-scale together
           with the sources classified in \citet{woods2011} (a) and
           YSO and background galaxy candidates by \citet{yso_gruendl} (b).
           The symbols are the same as in Figures~\ref{fig:2cdyso}a and b.
}
  \label{fig:cmdyso}
\end{figure}

In Figure~\ref{fig:cmdyso}a, 67 Archive sources classified in \citet{woods2011}
are highlighted as in Figure~\ref{fig:2cdyso}a.  The locations of each category source are
in agreement with the above discussion.  Red supergiants (purple circles)
are found to be concentrated on CM2.  
Post-AGB stars (squares) and planetary nebulae (stars) are distributed over a wide area in the CMD.
YSOs with the ice absorption are located at $[S11] < 7$\,mag, while those with the emission features
are distributed at $[S11] > 6.5$\,mag.

In Figure~\ref{fig:cmdyso}b, 
background galaxy and YSO candidates \citep{yso_gruendl}
are plotted on the CMD
together with the Archive sources.
Figure~\ref{fig:cmdyso} indicates that
YSO candidates are located around CM5 and
background galaxy candidates are distributed
over the brighter side of CM4.
\citet{yso_gruendl} exclude faint and red sources from the initial YSO candidates
with a criterion of $[8.0] > 14 - ([4.5] - [8.0])$
to avoid contamination from background galaxies.
Faint CM4 sources are likely to be excluded in this selection.
Therefore the entire CM4 population can be attributed to background galaxies.
These assignments of the features in the CMD are well in agreement with
those made in the CCD.  The locations of the YSOs with the 3\,$\mu$m water ice absorption (green circles) 
reported in \citet{shimonishi2010} and \citet{shimonishiPhD} are in agreement with those
classified by \citet{woods2011} in Figure~\ref{fig:cmdyso}a.

\section{Summary}\label{sec:summary}

We present the {\it AKARI} LMC Point Source Catalog,
which covers a core 10 deg$^2$ area of the LMC
with 5 photometric bands:
N3 (3.2\,$\mu$m), S7 (7\,$\mu$m), S11 (11\,$\mu$m), L15 (15\,$\mu$m) and L24 (24\,$\mu$m).
The 10-$\sigma$ limiting magnitudes are
17.9, 13.8, 12.4, 9.9, and 8.6 mag at N3, S7,
S11, L15 and L24, respectively.  They are comparable to those of the
{\it Spitzer} SAGE survey and the S11 and L15 data smoothly
fill the gap between the IRAC and MIPS bands.
The photometric accuracy is estimated to be better than 0.1 mag at N3
and 0.06--0.07 mag in the other bands.
The astrometric accuracy is 0$\farcs$3 at N3, S7
and S11, and 1$\farcs$0 at L15 and L24.

We provide a source Catalog and an Archive: 
The Archive contains all the detected sources, while the Catalog only includes sources
that have a counterpart in the {\it Spitzer} SAGE point source catalog.
Archive contains about 650,000, 140,000, 97,000, 43,000, and 52,000 sources
at 3.2, 7, 11, 15, and 24\,$\mu$m, respectively.

Based on the present catalog, we discuss the
LFs at 5 bands, a CCD, 
and a CMD of the N3, S7, and S11 data
for the sources towards the LMC.
Several noticeable features are recognized in the CCD and CMD, which
can be attributed to naked stars, dusty C-rich and O-rich AGB stars,
background galaxies and YSOs.  They are distributed at 
 distinct locations on the diagrams, indicating that the present data provide useful
 information on the classification of the sources towards the LMC.

\subsection*{Acknowledgement}
This work is based on observations with {\it AKARI}, a JAXA project with the participation of ESA. 
The authors thank all the members of the {\it AKARI} project and the members of the LMC survey
for their continuous help and encouragement.  They are grateful to F. Egusa for her contribution
to the IRC data reduction pipeline.  They also thank the {\it Spitzer} SAGE team for their
help.  This work is supported in part by Grants-in-Aid from
the Japan Society for the Promotion of Science (JSPS).

Facility: \facility{AKARI}

\begin{deluxetable}{ccccccccccccc}
\tabletypesize{\scriptsize}
\rotate

\tablecaption{Example of the Catalog Record  \label{tab:catsample}}
\tablewidth{0pt}
\tablehead{
\colhead{ID} &  \colhead{name} &  \colhead{ID2} 
&  \colhead{R.A.} &  \colhead{DEC.} &  \colhead{pos-flag}
 &  \colhead{N3x}     &  \colhead{N3y}    &  \colhead{N3mag}  &  \colhead{N3me}
 &  \colhead{N3me2}
 &  \colhead{N3chi}   &  \colhead{N3shrp} \\
  \colhead{(1)}  &  \colhead{(2)}  &  \colhead{(3)}  &  \colhead{(4)}  &  \colhead{(5)}  
  &  \colhead{(6)}  &  \colhead{(7)}  &  \colhead{(8)}  &  \colhead{(9)}  &  \colhead{(10)}
    & \colhead{(11)} &  \colhead{(12)} & \colhead{(13)}}
\startdata
    56000001 & AKARI-LMCC & J050737.46-670810.9
    & 76.906067 & -67.136353 & 3
                & 999.999 & 999.999 & 99.999 & 99.999 & 99.999 & 99.999
    & 99.999\\
    56000002 & AKARI-LMCC & J050737.51-670831.8
    & 76.906273 & -67.142159 & 3
                & 999.999 & 999.999 & 99.999 & 99.999 & 99.999 & 99.999
    & 99.999\\
    56000003 & AKARI-LMCC & J050738.23-670640.8
    & 76.909279 & -67.111336 & 3
                & 999.999 & 999.999 & 99.999 & 99.999 & 99.999 & 99.999
    & 99.999 \\
    56000004 & AKARI-LMCC & J050745.94-671413.4
    & 76.941399 & -67.237053 & 3
                & 999.999 & 999.999 & 99.999 & 99.999 & 99.999 & 99.999
    & 99.999 \\
    56000005 & AKARI-LMCC & J050745.96-670811.4
    & 76.941513 & -67.136513 & 3
                & 999.999 & 999.999 & 99.999 & 99.999 & 99.999 & 99.999
    & 99.999 \\
    56000006 & AKARI-LMCC & J050747.82-670821.3
    & 76.949265 & -67.139259 & 3
                & 999.999 & 999.999 & 99.999 & 99.999 & 99.999 & 99.999
    & 99.999 \\
    56000007 & AKARI-LMCC & J050749.56-671211.2
    & 76.956482 & -67.203102 & 3
                & 999.999 & 999.999 & 99.999 & 99.999 & 99.999 & 99.999
    & 99.999 \\
    56000008 & AKARI-LMCC & J050754.92-665746.1
    & 76.978851 & -66.962814 & 3
                & 999.999 & 999.999 & 99.999 & 99.999 & 99.999 & 99.999
    & 99.999 \\
    56000009 & AKARI-LMCC & J050758.73-671218.7
    & 76.994720 & -67.205200 & 3
                & 999.999 & 999.999 & 99.999 & 99.999 & 99.999 & 99.999
    & 99.999 \\
    56000010 & AKARI-LMCC & J050759.70-670035.7
    & 76.998749 & -67.009911 & 3
                & 999.999 & 999.999 & 99.999 & 99.999 & 99.999 & 99.999
    & 99.999  \\
     ...  & ... & ...  & ... & ... & ...  & ... & ... & ... & ... & ... & ... & ...  \\   
    \enddata

\end{deluxetable}

\setcounter{table}{0}

\begin{deluxetable}{cccccccccccccc}
\tabletypesize{\scriptsize}
\rotate

\tablecaption{continued}
\tablewidth{0pt}
\tablehead{ \colhead{N3date} & \colhead{N3pid}
 & \colhead{N3flag}
 &  \colhead{S7Nmat}  &  \colhead{S7dis}    &  \colhead{S7x}     &  \colhead{S7y}    
 &  \colhead{S7mag}  &  \colhead{S7me} & \colhead{S7me2} 
 & \colhead{S7chi}   &  \colhead{S7shrp}  &  \colhead{S7date} 
 \\
  \colhead{(14)} &  \colhead{(15)} 
    &  \colhead{(16)} &  \colhead{(17)} &  \colhead{(18)} &  \colhead{(19)} &  \colhead{(20)}
    & \colhead{(21)} &  \colhead{(22)} &  \colhead{(23)} &  \colhead{(24)} &  \colhead{(25)} 
    & \colhead{(26)}}
\startdata
    9999-99-99T99:99:99 & 9999999.9 & 9999999
    & 0 & 99.99 & 999.999 & 999.999 & 99.999 & 99.999 & 99.999 & 99.999
    & 99.999 & 9999-99-99T99:99:99\\
    9999-99-99T99:99:99 & 9999999.9 & 9999999
    & 0 & 99.99 & 999.999 & 999.999 & 99.999 & 99.999 & 99.999 & 99.999
    & 99.999 & 9999-99-99T99:99:99\\
    9999-99-99T99:99:99 & 9999999.9 & 9999999
    & 0 & 99.99 & 999.999 & 999.999 & 99.999 & 99.999 & 99.999 & 99.999
    & 99.999 & 9999-99-99T99:99:99\\
    9999-99-99T99:99:99 & 9999999.9 & 9999999
    & 0 & 99.99 & 999.999 & 999.999 & 99.999 & 99.999 & 99.999 & 99.999
    & 99.999 & 9999-99-99T99:99:99\\
    9999-99-99T99:99:99 & 9999999.9 & 9999999
    & 0 & 99.99 & 999.999 & 999.999 & 99.999 & 99.999 & 99.999 & 99.999
    & 99.999 & 9999-99-99T99:99:99\\
    9999-99-99T99:99:99 & 9999999.9 & 9999999
    & 0 & 99.99 & 999.999 & 999.999 & 99.999 & 99.999 & 99.999 & 99.999
    & 99.999 & 9999-99-99T99:99:99\\
    9999-99-99T99:99:99 & 9999999.9 & 9999999
    & 0 & 99.99 & 999.999 & 999.999 & 99.999 & 99.999 & 99.999 & 99.999
    & 99.999 & 9999-99-99T99:99:99\\
    9999-99-99T99:99:99 & 9999999.9 & 9999999
    & 0 & 99.99 & 999.999 & 999.999 & 99.999 & 99.999 & 99.999 & 99.999
    & 99.999 & 9999-99-99T99:99:99\\
    9999-99-99T99:99:99 & 9999999.9 & 9999999
    & 0 & 99.99 & 999.999 & 999.999 & 99.999 & 99.999 & 99.999 & 99.999
    & 99.999 & 9999-99-99T99:99:99\\
    9999-99-99T99:99:99 & 9999999.9 & 9999999
    & 0 & 99.99 & 999.999 & 999.999 & 99.999 & 99.999 & 99.999 & 99.999
    & 99.999 & 9999-99-99T99:99:99\\   
         ...  & ... & ...  & ... & ... & ...  & ... & ... & ... & ... & ... & ... & ...  \\   
     \enddata

\end{deluxetable}

\setcounter{table}{0}

\begin{deluxetable}{ccccccccccccc}
\tabletypesize{\scriptsize}
\rotate

\tablecaption{continued}
\tablewidth{0pt}
\tablehead{  \colhead{S7pid}
 & \colhead{S7flag} & \colhead{S11Nmat} & \colhead{S11dis}   &  \colhead{S11x}    
 &  \colhead{S11y}   &  \colhead{S11mag} &  \colhead{S11me} &  \colhead{S11me2}
 &  \colhead{S11chi}  &  \colhead{S11shrp} &  \colhead{S11date} & \colhead{S11pid}
  \\
    \colhead{(27)} &  \colhead{(28)} &  \colhead{(29)} &  \colhead{(30)}
    & \colhead{(31)} &  \colhead{(32)} &  \colhead{(33)} &  \colhead{(34)} & \colhead{(35)} 
    &  \colhead{(36)} &  \colhead{(37)} &  \colhead{(38)} &  \colhead{(39)}
    }
\startdata
    9999999.9 & 9999999 & 0 & 99.99 & 999.999 & 999.999 & 99.999 & 99.999 & 99.999 & 99.999
    & 99.999 & 9999-99-99T99:99:99 & 9999999.9\\
    9999999.9 & 9999999 & 0 & 99.99 & 999.999 & 999.999 & 99.999 & 99.999 & 99.999 & 99.999
    & 99.999 & 9999-99-99T99:99:99 & 9999999.9\\
    9999999.9 & 9999999 & 0 & 99.99 & 999.999 & 999.999 & 99.999 & 99.999 & 99.999 & 99.999
    & 99.999 & 9999-99-99T99:99:99 & 9999999.9\\
    9999999.9 & 9999999 & 0 & 99.99 & 999.999 & 999.999 & 99.999 & 99.999 & 99.999 & 99.999
    & 99.999 & 9999-99-99T99:99:99 & 9999999.9\\
    9999999.9 & 9999999 & 0 & 99.99 & 999.999 & 999.999 & 99.999 & 99.999 & 99.999 & 99.999
    & 99.999 & 9999-99-99T99:99:99 & 9999999.9\\
    9999999.9 & 9999999 & 0 & 99.99 & 999.999 & 999.999 & 99.999 & 99.999 & 99.999 & 99.999
    & 99.999 & 9999-99-99T99:99:99 & 9999999.9\\
    9999999.9 & 9999999 & 0 & 99.99 & 999.999 & 999.999 & 99.999 & 99.999 & 99.999 & 99.999
    & 99.999 & 9999-99-99T99:99:99 & 9999999.9\\
    9999999.9 & 9999999 & 0 & 99.99 & 999.999 & 999.999 & 99.999 & 99.999 & 99.999 & 99.999
    & 99.999 & 9999-99-99T99:99:99 & 9999999.9\\
    9999999.9 & 9999999 & 0 & 99.99 & 999.999 & 999.999 & 99.999 & 99.999 & 99.999 & 99.999
    & 99.999 & 9999-99-99T99:99:99 & 9999999.9 \\
    9999999.9 & 9999999 & 0 & 99.99 & 999.999 & 999.999 & 99.999 & 99.999 & 99.999 & 99.999
    & 99.999 & 9999-99-99T99:99:99 & 9999999.9 \\    
         ...  & ... & ...  & ... & ... & ...  & ... & ... & ... & ... & ... & ... & ...  \\   
    \enddata

\end{deluxetable}

\setcounter{table}{0}

\begin{deluxetable}{ccccccccccccc}
\tabletypesize{\scriptsize}
\rotate

\tablecaption{continued}
\tablewidth{0pt}
\tablehead{  \colhead{S11flag}
 &  \colhead{L15Nmat} &  \colhead{L15dis}   &  \colhead{L15x}  
 & \colhead{L15y}  &  \colhead{L15mag}
 & \colhead{L15me}    & \colhead{L15me2}   &  \colhead{L15chi}  &  \colhead{L15shrp}
 &  \colhead{L15date} & \colhead{L15pid} 
 &  \colhead{L15flag} \\
     \colhead{(40)}
    & \colhead{(41)} & \colhead{(42)}  &  \colhead{(43)} &  \colhead{(44)} &  \colhead{(45)} 
    &  \colhead{(46)} &  \colhead{(47)} &  \colhead{(48)} &  \colhead{(49)} & \colhead{(50)}
    & \colhead{(51)} & \colhead{(52)}}
\startdata
    9999999
    & 0 & 99.99 & 206.406 &  18.121 & 10.662 &  0.207 &  0.209 &  0.075
    &  0.052 & 2007-05-28T11:25:56 & 2213033.1 & 1100000
    \\
    9999999
    & 0 & 99.99 & 197.784 &  16.530 &  9.297 &  0.072 &  0.073 &  0.107
    & -0.085 & 2007-05-28T11:25:56 & 2213033.1 & 1100000
    \\
    9999999
    & 0 & 99.99 & 243.089 &  27.315 & 10.074 &  0.143 &  0.144 &  0.114
    & -0.189 & 2007-05-28T11:25:56 & 2213033.1 & 1100000
    \\
    9999999
    & 0 & 99.99 &  53.203 &   8.753 & 10.766 &  0.251 &  0.253 &  0.100
    &  0.036 & 2007-05-28T11:25:56 & 2213033.1 & 1100000
    \\
    9999999
    & 0 & 99.99 & 202.099 &  38.444 &  8.544 &  0.052 &  0.053 &  0.175
    & -0.077 & 2007-05-28T11:25:56 & 2213033.1 & 1100000
    \\
    9999999
    & 0 & 99.99 & 197.140 &  42.089 & 10.436 &  0.197 &  0.199 &  0.104
    & -0.161 & 2007-05-28T11:25:56 & 2213033.1 & 1100000
    \\
    9999999
    & 0 & 99.99 & 101.755 &  27.399 & 10.834 &  0.255 &  0.257 &  0.090
    & -0.152 & 2007-05-28T11:25:56 & 2213033.1 & 1100000
    \\
    9999999
    & 0 & 99.99 & 237.879 &  28.323 &  8.954 &  0.062 &  0.063 &   0.131
    &   0.128 & 2007-06-02T23:59:44 & 2213035.1 & 1100000
    \\
    9999999
    & 0 & 99.99 &  94.249 &  48.691 &  9.652 &  0.097 &  0.098 &  0.108
    &  0.015 & 2007-05-28T11:25:56 & 2213033.1 & 1100000
    \\
   9999999
    & 0 & 99.99 & 165.741 &  32.491 & 10.831 &  0.236 &  0.239 &  0.067
    &  0.063 & 2007-06-02T23:59:44 & 2213035.1 & 1100000
    \\    
         ...  & ... & ...  & ... & ... & ...  & ... & ... & ... & ... & ... & ... & ...  \\   
    \enddata
    
\end{deluxetable}
    
\setcounter{table}{0}

\begin{deluxetable}{cccccccccccc}
\tabletypesize{\scriptsize}
\rotate

\tablecaption{continued}
\tablewidth{0pt}
\tablehead{   \colhead{L24Nmat} &  \colhead{L24dis}   
 &  \colhead{L24x}    &  \colhead{L24y}   &  \colhead{L24mag} &  \colhead{L24me}
 &  \colhead{L24me2} &  \colhead{L24chi}  &  \colhead{L24shrp} &
 \colhead{L24date} &   \colhead{L24pid} & \colhead{L24flag}\\
     \colhead{(53)}
    & \colhead{(54)} & \colhead{(55)} & \colhead{(56)}
    & \colhead{(57)} &  \colhead{(58)} &  \colhead{(59)} &  \colhead{(60)} &  \colhead{(61)} 
    &  \colhead{(62)} &  \colhead{(63)} &  \colhead{(64)} }
\startdata
    0 & 99.99 & 999.999 & 999.999 & 99.999 & 99.999 & 99.999 & 99.999
    & 99.999 & 9999-99-99T99:99:99 & 9999999.9 & 9999999\\
    1 &  0.27 & 193.778 &  11.330 &  9.226 &  0.225 &  0.229 &  0.154
    & -0.260 & 2007-05-28T11:27:37 & 2213033.1 & 1100000\\
    0 & 99.99 & 999.999 & 999.999 & 99.999 & 99.999 & 99.999 & 99.999
    & 99.999 & 9999-99-99T99:99:99 & 9999999.9 & 9999999\\
    0 & 99.99 & 999.999 & 999.999 & 99.999 & 99.999 & 99.999 & 99.999
    & 99.999 & 9999-99-99T99:99:99 & 9999999.9 & 9999999\\
    1 &  1.23 & 197.763 &  33.093 &  7.008 &  0.035 &  0.038 &  0.161
    & -0.020 & 2007-05-28T11:27:37 & 2213033.1 & 1100000\\
    0 & 99.99 & 999.999 & 999.999 & 99.999 & 99.999 & 99.999 & 99.999
    & 99.999 & 9999-99-99T99:99:99 & 9999999.9 & 9999999\\
    0 & 99.99 & 999.999 & 999.999 & 99.999 & 99.999 & 99.999 & 99.999
    & 99.999 & 9999-99-99T99:99:99 & 9999999.9 & 9999999\\
    1 &  1.10 & 233.541 &  22.642 &  7.041 &  0.029 &  0.034 &   0.096
    &  -0.061 & 2007-06-03T00:01:26 & 2213035.1 & 1100000\\
    1 &  0.36 &  90.170 &  43.412 &  8.157 &  0.078 &  0.081 &  0.122
    &  0.125 & 2007-05-28T11:27:37 & 2213033.1 & 1100000\\
   0 & 99.99 & 999.999 & 999.999 & 99.999 & 99.999 & 99.999 & 99.999
    & 99.999 & 9999-99-99T99:99:99 & 9999999.9 & 9999999\\    
         ...  & ... & ...  & ... & ... & ...  & ... & ... & ... & ... & ... & ...  \\   
    \enddata

\end{deluxetable}

\begin{deluxetable}{cccccccl}

\tabletypesize{\scriptsize}
\rotate
  \tablecaption{Catalog Content \label{tab:content}}
\tablewidth{0pt}
\tablehead{
   \colhead{Item} & \colhead{format} & \multicolumn{5}{c}{Record position (byte)}
      & \colhead{Content}} 
\startdata
 \multicolumn{2}{l}{Common record}\\
   ID1     & A8    & 1:8   & & & & & source ID1:  56****** for Catalog and 51****** for Archive\\
   Cname   & A10   & 10:19 & & & & & AKARI-LMCC for Catalog and AKARI-LMCA for Archive\\ 
   ID2     & A19   & 21:39 & & & & & source ID2 in Jhhmmss.ss-ddmmss.s \\
   R.A.    & F10.6 & 41:50 & & & & & right ascension (J2000.0) in degree\\
   Decl    & F10.6 & 52:61 & & & & & declination (J2000.0) in degree\\
   eq-flag & A1    & 63:63 & & & & & band at which the position is derived\\
\\
 \multicolumn{2}{l}{Band specific} &  N3 & S7 & S11 & L15 & L24\\
   x   & F7.3 & 65:71  & 165:171 & 265:271 & 365:371 & 465:471 & pixel X-coordinate\\
   y   & F7.3 & 73:79  & 173:179 & 273:279 & 373:379 & 473:479 & pixel Y-coordinate\\
   mag & F6.3 & 81:86  & 181:186 & 281:286 & 381:386 & 481:486 & magnitude in the Vega system uncorrected for extinction\\
   me1 & F6.3 & 88:93  & 188:193 & 288:293 & 388:393 & 488:493 & photometric uncertainty from ALLSTAR\\
   me2 & F6.3 & 95:100 & 195:200 & 295:300 & 395:400 & 495:500 & total photometric 
   uncertainty\tablenotemark{a}
   \\
   $\chi$ & F7.3 & 102:108 & 202:208 & 302:308 & 402:408 & 502:508 & PSF fit 
   parameter\tablenotemark{b}\\
   sharpness & F7.3 & 110:116 & 210:216 & 310:316 & 410:416 & 510:516 & PSF fit 
   parameter\tablenotemark{c}\\
   date & A19 & 118:136 & 218:236 & 318:336 & 418:436 & 518:536 & observation date in UT\\
   PID  & A9  & 138:146 & 238:246 & 338:346 & 438:446 & 538:546 & pointing ID\\
   flag & A7  & 148:154 & 248:254 & 348:354 & 448:454 & 548:554 & 7-digit code for the object (see
    text and Table~\ref{tab:flag})\\
   Nmatch   & I2   & --- & 156:157 & 256:257 & 356:357 & 456:457 & number of matched sources (see text for details)\\
   distance & F5.2 & --- & 159:163 & 259:263 & 359:363 & 459:463 & distance from the nearest matched source in arcsec (see text for details)\\
\enddata
 \tablenotetext{a}{Including errors in the aperture correction, read-out noise, and conversion factor.
                   }
 \tablenotetext{b}{The parameter $\chi$ increases with the difference
in shape between the source and the PSF-profile. The values 99.999 is assigned when
the aperture photometry is employed.}
 \tablenotetext{c}{A positive (negative) sharpness indicates that
the source is more (less) extended than the PSF-profile,
suggesting that the source may be 
a galaxy, an unresolved multi-star system, a
cosmic ray, or image defect. The values 99.999 is assigned when
the aperture photometry is employed.
}

\end{deluxetable}

\begin{deluxetable}{lcccccc}

\tabletypesize{\scriptsize}
\rotate
  \tablecaption{Observation Parameters \label{tab:survey}}
\tablewidth{0pt}
\tablehead{
   \colhead{Properties} & \multicolumn{6}{c}{IRC bands}\\
      & \colhead{NP} & \colhead{N3} & \colhead{S7} & \colhead{S11} 
      & \colhead{L15} & {L24}}
\startdata
    Channel & NIR & NIR & MIR-S & MIR-S & MIR-L & MIR-L\\
    Bandpass ($\mu$m) &
    1.8--5.5 & 2.7--3.8 & 5.9--8.4 & 8.5--13.1 & 12.6--19.4 & 20.3--26.5\\
    Reference wavelength ($\mu$m) &
    --- & 3.2 & 7.0 & 11.0 & 15.0 & 24.0\\
    Pixel field of view ($\arcsec$\,pixel$^{-1}$) &
    --- & 1.446 & 2.340 & 2.340 & 2.384 & 2.384\\
    Dispersion ($\mu$m\,pixel$^{-1}$) &
    0.06 at 3.5$\mu$m & ---& --- & --- & --- & ---\\
    Exposure time: long exposure\tablenotemark{a} (s) &
    133 & 133 & 147 & 147 & 147 & 147\\
    Exposure time: short exposure\tablenotemark{a} (s) &
    14.0 & 14.0 & 1.75 & 1.75 & 1.75 & 1.75\\
    10-$\sigma$ detection limit\tablenotemark{b} (mJy) &
    - & 0.024 & 0.226 & 0.419 & 1.758 & 2.921\\
    10-$\sigma$ detection limit\tablenotemark{b} (mag) &
    - & 17.9 & 13.8 & 12.4 &  9.9 &  8.6\\
    Saturation limit\tablenotemark{b}$^,$\tablenotemark{c} (mJy) &
    12500 at 3.5$\mu$m & 250 & 1800 & 1800 & 2500 & 23000\\
    Saturation limit\tablenotemark{b}$^,$\tablenotemark{c} (mag) &
    3.6 at 3.5$\mu$m & 7.8 & 4.0 & 3.3 & 2.0 & -1.1\\
    Zero magnitude flux density\tablenotemark{d} (Jy) &
    - & 343 & 75.0 & 38.3 & 16.0 & 8.0\\
    Number of sources in Archive &
    - & 720872 & 141617 &  97226 &  42974 &  52266\\
    Number of sources in Catalog &
    - & 651294 &  89610 &  48525 &  16712 &   6856\\
\enddata
 \tablenotetext{a}{These are the nominal values.
                   The total exposure time per pixel depends on
                   the co-adding process.
                   }
 \tablenotetext{b}{For point sources.}
 \tablenotetext{c}{Numbers are taken from ASTRO-F Observer's Manual
                   version 3.2.2.}
  \tablenotetext{d}{From \cite{tanabe}.}
\end{deluxetable}

\begin{deluxetable}{ll}
\tablewidth{0pt}
  \tablecaption{Description of Flags  \label{tab:flag}}
\tablehead{
    \colhead{Flag} & \colhead{Description} }
\startdata
    exposure        & 0: short; 1: long.\\
    photometry      & 0: aperture photometry; 1: PSF fitting photometry.\\
    multiple        & Number of the nearby sources.\\
    mux bleed       & 1: contamination by mux bleed; only for N3.\\
    column pulldown & 1: contaminated by column pulldown;  only for N3.\\
    artifact         & 1: contaminated by artifact; only for S7 and S11.\\
    \enddata
\end{deluxetable}

\vspace{50pt}

\begin{deluxetable}{ccrcr}
\tablewidth{0pt}
  \tablecaption{Internal Dispersion of
           Photometry and Astrometry  \label{tab:accuracy}
}
\tablehead{
    \colhead{band} & \colhead{$\sigma$ [mag]} & \colhead{$\sigma$ [$\arcsec$]} 
    & \colhead{S/N} &
    \colhead{N$_\mathrm{source}$}}
    \startdata
    N3  & 0.091 & 0.607 & $\ge$30 & 10805 \\
    S7  & 0.071 & 0.413 & $\ge$20 &  3500 \\
    S11 & 0.058 & 0.405 & $\ge$20 &  1314 \\
    L15 & 0.069 & 0.962 & $\ge$20 &   497 \\
    L24 & 0.063 & 1.031 & $\ge$20 &   455 \\
    \enddata
\end{deluxetable}

\begin{deluxetable}{ccccccc}
\tablewidth{0pt}
  \tablecaption{Astrometric Accuracy  \label{tab:astrometry}}
\tablehead{
    \colhead{catalog} & 
    \multicolumn{2}{c}{offset} &
    \multicolumn{3}{c}{standard deviation} &
    \colhead{accuracy} \\
     & \colhead{$\Delta \alpha$ }& \colhead{$\Delta \delta$}
     & \colhead{$\Delta \alpha$} & \colhead{$\Delta \delta$} & \colhead{r}
%     & \\
     }
\startdata
    2MASS-PSC & -0$\farcs$064 & 0$\farcs$056 & 0$\farcs$220 & 0$\farcs$307
              & 0$\farcs$393 & $\sim$0$\farcs$07\tablenotemark{a}\\
    SAGE-PSC  & -0$\farcs$058 & 0$\farcs$057 & 0$\farcs$293 & 0$\farcs$421
              & 0$\farcs$509 & $\sim$0$\farcs$3\tablenotemark{b}\\

\enddata
 \tablenotetext{a}{\cite{2mass}.}
  \tablenotetext{b}{\cite{sage_report}.}

\end{deluxetable}

 \begin{deluxetable}{ccrrr}
 \tablecaption{10-$\sigma$ Limiting Magnitude and
           90\% Completeness Limit \label{tab:limit}}
\tablewidth{0pt}
\tablehead{
    \colhead{band} & 
   \colhead{10-$\sigma$ limiting} &
    \multicolumn{3}{c}{90\% completeness limit}\\
    & \colhead{magnitude} & 
    \colhead{high (mag)\tablenotemark{a}}   &
    \colhead{medium (mag)\tablenotemark{b}} &
    \colhead{blank (mag)\tablenotemark{c}} }
    \startdata
    N3  & 17.9 & 13.46 $\pm$ 0.13 & 14.58 $\pm$ 0.08 & 15.66 $\pm$ 0.08 \\
    S7  & 13.8 & 10.73 $\pm$ 0.41 & 13.40 $\pm$ 0.87 & 14.06 $\pm$ 0.18 \\
    S11 & 12.4 &  9.12 $\pm$ 0.73 & 12.62 $\pm$ 0.08 & 12.84 $\pm$ 0.05 \\
    L15 &  9.9 &  7.70 $\pm$ 1.21 & 10.68 $\pm$ 0.29 & 11.02 $\pm$ 0.13 \\
    L24 &  8.6 &  5.88 $\pm$ 1.12 &  9.32 $\pm$ 0.25 &  9.50 $\pm$ 0.07 \\
    \enddata
       \tablenotetext{a}{High, medium, and blank correspond to the sky background level 
       (see text).}
\end{deluxetable}

\begin{deluxetable}{crrrrc}
 \tablecaption{Band Merging Results\tablenotemark{a} 
            \label{tab:match}}
\tablewidth{0pt}
\tablehead{
    \colhead{band} & \multicolumn{2}{c}{Archive}     & 
           \multicolumn{2}{c}{Catalog}    \\
         & \colhead{N$_\mathrm{all}$\tablenotemark{b}}   & 
           \colhead{N$_\mathrm{match}$\tablenotemark{c}} & 
           \colhead{N$_\mathrm{all}$\tablenotemark{b}}   & 
           \colhead{N$_\mathrm{match}$\tablenotemark{c}} & Matching band}
    \startdata
    S7   &  141617 &  101425 &  89610  &  85791 &  N3  \\
    S11  &   97226 &  67254 &  48525  &  43355 &  S7  \\
    L15  &   42974 &  27141 &  16712  &  12868 &  S11 \\
    L24  &   52266 & 23062 &   6856  &   5545 &  L15 \\
    \enddata
    \tablenotetext{a}{For merging strategy, see text.}
    \tablenotetext{b}{Total number of sources.} 
   \tablenotetext{c}{Number of sources with a counterpart
           in the neighboring shorter wavelength band.}
\end{deluxetable}

%
%Sensitivities
%

\end{document}